\documentclass[comsoc,doublecolumn,final]{IEEEtran}

\usepackage[utf8]{inputenc}
\usepackage[T1]{fontenc}
\usepackage[cmintegrals]{newtxmath}

\usepackage{cite}
\usepackage{microtype}
\ifCLASSINFOpdf
\usepackage[pdftex]{graphicx}
\DeclareGraphicsExtensions{.pdf,.jpeg,.png,.JPG}
\else
\usepackage[dvips]{graphicx}
\DeclareGraphicsExtensions{.eps}
\fi
\graphicspath{{./graphics/}}
\usepackage{mathtools}
\usepackage[caption=false,font=footnotesize]{subfig}
\usepackage{url}
\usepackage{booktabs}
\usepackage{siunitx}
\usepackage{xcolor}

\usepackage{iitem}
\usepackage{bm}

\newcommand{\FAU}{Friedrich-Alexander-Universit\"at Erlangen-N\"urnberg (FAU)}

\newcommand{\EFI}{Emerging Fields Initiative (EFI)}
\newcommand{\BMBF}{German Federal Ministry of Eduction and Research (BMBF)}
\newcommand{\vel}{u}
\newcommand{\vmax}{u_{0}}
\newcommand{\veff}{u_\mathrm{eff}}
\newcommand{\visc}{\nu}

\newcommand{\tpeak}{t_\mathrm{peak}}
\newcommand{\hpeak}{h_\mathrm{peak}}
\newcommand{\ipeak}{i_\mathrm{peak}}
\newcommand{\istart}{i_\mathrm{start}}
\newcommand{\tstart}{t_0}
\newcommand{\Ki}{K_\mathrm{i}}
\newcommand{\Ktrain}{{K_\mathrm{t}}}
\newcommand{\atrain}{\bm{a}_\mathrm{t}}
\newcommand{\rtrain}{\bm{r}_\mathrm{t}}
\newcommand{\ainfo}{\bm{a}_\mathrm{i}}

\newcommand{\Reyn}{\mathrm{Re}}
\newcommand{\de}{\,\mathrm{d}}
\newcommand{\Qb}{Q_\mathrm{b}}
\newcommand{\Qp}{Q_\mathrm{p}}
\newcommand{\Vi}{V_\mathrm{i}}

\newcommand{\cin}{c_\mathrm{i}}
\newcommand{\chiref}{\chi_\mathrm{ref}}

\newcommand{\scal}{\gamma}
\newcommand{\myscal}{c_d}

\DeclareMathOperator{\rect}{rect}

\newcommand{\dt}{\Delta t}
\DeclareMathOperator*{\argmin}{arg\,min}

\newcommand{\reals}{\mathbb{R}}

\newcommand{\sampoff}{I_\mathrm{off}}
\newcommand{\kB}{k_\textsc{b}}
\newcommand{\Tm}{T_\mathrm{m}}
\newcommand{\Rp}{R_\mathrm{p}}
\newcommand{\dispfac}{\alpha_\mathrm{D}}
\newcommand{\chardist}{a_\mathrm{c}}

\newcommand{\chim}{\chi_\mathrm{m}}

\newcommand{\changed}[2][]{{#2}}

\title{Experimental System for Molecular Communication in Pipe Flow With Magnetic Nanoparticles}

\author{%
    Wayan Wicke,~\IEEEmembership{Student Member,~IEEE,}
    Harald Unterweger,
    Jens Kirchner,~\IEEEmembership{Senior Member,~IEEE,}
    Lukas Brand,
    Arman Ahmadzadeh,~\IEEEmembership{Member,~IEEE,}
    Doaa Ahmed,
    Vahid Jamali,~\IEEEmembership{Member,~IEEE,}
    Christoph Alexiou,
    Georg Fischer,~\IEEEmembership{Senior Member,~IEEE,}
    and Robert Schober,~\IEEEmembership{Fellow,~IEEE,}%
    \thanks{%
        This work was supported in part by the \EFI{} of the \FAU{}, the STAEDTLER-Stiftung, and the \BMBF{}, project MAMOKO.
        This  work  was  presented  in  part  at IEEE SPAWC 2018 \cite{unterweger_experimental_2018}.
        \textit{(Corresponding author: Wayan Wicke.)}
    }%
    \thanks{%
        W.~Wicke, L.~Brand, A.~Ahmadzadeh, V.~Jamali, and R.~Schober are with the Institute for Digital Communications, Friedrich-Alexander-Universit\"at Erlangen-N\"urnberg (FAU), 91058 Erlangen, Germany (e-mail: wayan.wicke@fau.de; lukas.brand@fau.de; arman.ahmadzadeh@fau.de; vahid.jamali@fau.de; robert.schober@fau.de).
    }%
    \thanks{%
        H.~Unterweger and C.~Alexiou are with the Section for Experimental Oncology and Nanomedicine (SEON), Universit\"atsklinikum Erlangen, 91012 Erlangen, Germany  (email: harald.unterweger@uk-erlangen.de; christoph.alexiou@uk-erlangen.de).%
    }%
    \thanks{%
    J.~Kirchner, D.~Ahmed, and G.~Fischer are with Institute for Electronics Engineering, Friedrich-Alexander-Universit\"at Erlangen-N\"urnberg (FAU), 91058 Erlangen, Germany (jens.kirchner@fau.de; doaa.ahmed@fau.de; georg.fischer@fau.de).
    }%
}

\begin{document}
%

\maketitle
\IEEEpeerreviewmaketitle

\begin{abstract}
    In the emerging field of molecular communication (MC), testbeds are needed to validate theoretical concepts, motivate applications, and guide further modeling efforts.
    \changed{To this end, this paper presents a flexible and extendable in-vessel testbed for flow-based macroscopic MC, abstractly modeling, e.g., a part of a chemical reactor or a blood vessel.
    Signaling is based on injecting non-reactive superparamagnetic iron oxide nanoparticles (SPIONs) dispersed in an aqueous suspension into a tube with background flow.
    A commercial magnetic susceptometer is used for non-intrusive downstream signal reception.
    To shed light on the operation of the testbed, we identify the physical mechanisms governing the transmission, propagation, and reception of the information-carrying SPIONs.
    Moreover, to facilitate system design, we propose a closed-form parametric expression for the end-to-end channel impulse response (CIR).}
    The proposed CIR model is shown to consistently capture the experimentally observed distance-dependent impulse response peak heights and peak decays for transmission distances from \SIrange{5}{40}{\centi\meter}.
    \changed{Moreover, to validate our testbed, reliable communication is demonstrated based on experimental data for model-agnostic and model-based detection methods.}
\end{abstract}


\section{Introduction}
\label{sec:introduction}
Molecular communication (MC) employs molecules as information carriers necessitating new models and experimental tools compared to conventional electromagnetic wave based communication \cite{nakano_molecular_2013}.
The growing interest in this research area is due to its revolutionary applications in environments unsuited for electromagnetic waves such as biological environments, e.g., in the human body or within bacterial cultures \cite{grebenstein_biological_2019}, environments unfavorable due to propagation losses, e.g., liquid-filled pipes, and environments with explosive gases, e.g., fuel pipes \cite{nakano_molecular_2013,akyildiz_internet_2015,haselmayr_integration_2019}.
Motivated by these applications, a significant body of theoretical work has been developed, see \cite{farsad_comprehensive_2016,jamali_channel_2019,kuscu_transmitter_2019} for surveys of the current literature.
Moreover, for practical demonstration and to gain more insight regarding relevant physical phenomena, several MC testbeds have been proposed, see \cite[Section V]{jamali_channel_2019} for an overview of recent artificial and biological MC testbeds.

These experimental systems can be categorized as either air-based \cite{farsad_tabletop_2013,giannoukos_molecular_2017,shakya_correlated_2018,damrath_investigation_2021} or liquid-based \cite{farsad_novel_2017,khaloopour_experimental_2019,tuccitto_reactive_2018,atthanayake_experimental_2018,koo_deep_2020,kuscu_graphenebased_2020} depending on the physical communication medium.
Air-based systems have been developed for open space transmission \cite{farsad_novel_2017} as well as for closed air ducts \cite{giannoukos_molecular_2017,shakya_correlated_2018,damrath_investigation_2021}, which offer a directed information transfer at the expense of the required infrastructure.
Liquid-based experimental MC systems usually require vessels and exist in different size scales ranging from microfluidics \cite{kuscu_graphenebased_2020,koo_deep_2020}, to small pipes \cite{farsad_novel_2017,tuccitto_reactive_2018}, to larger ducts \cite{khaloopour_experimental_2019,atthanayake_experimental_2018}.

In this paper, we study a liquid-based MC system composed of small pipes, i.e., the environment is bounded, flow-driven, and fluid, \changed{similar to} blood vessels.
Experimental systems studying such environments have been reported in \cite{farsad_novel_2017,tuccitto_reactive_2018}.
The system in \cite{farsad_novel_2017} is based on either injecting an acid or a base into water and the detection of the medium's pH level.
The system in \cite{tuccitto_reactive_2018} is similar to the one in \cite{farsad_novel_2017} in that it is based on in-vessel chemical reactions but it employs optical detection.
However, those systems inherently rely on chemical reactions which complicates their analysis (see e.g., \cite{jamali_diffusive_2018}) and limits their applicability because many applications require passive signaling to avoid possible interference with biological processes.
Moreover, for detection, the system in \cite{farsad_novel_2017} requires direct access to the liquid and the system in \cite{tuccitto_reactive_2018} requires an optically transparent tubing.

In this paper, we present a new testbed with the focus on studying flow-driven transport in simple cylindrical tube systems.
To this end, it is crucial to select appropriate signaling molecules or particles \cite{levenspiel_chemical_1999,soldner_survey_2020}.
These signaling particles should ideally possess the following properties which the information carriers used in \cite{farsad_novel_2017,khaloopour_experimental_2019,tuccitto_reactive_2018} do not have:
1) The particles should be chemically stable for safe and long-term use, i.e., they should not agglomerate and not interact with other components of the testbed, such as the respective medium.
2) A sensitive and non-intrusive detection mechanism is required since, depending on the application, physical access to the tubing may not be feasible or practical. 
3) The particles should ideally be tunable for different application needs, e.g., in their size, and have an established production mechanism for cost-efficiency.
\changed{These requirements cannot be met by many potential chemical information carriers which typically rely on chemical reactions for detection, e.g., the detection of acids and bases with a pH-meter \cite{farsad_novel_2017}.
An attractive option in this regard are dye colored nanoparticles which can be detected optically.
However, this would necessitate a transparent tubing.
A promising alternative type of artificial particles that satisfy all of the above requirements and are already well-established in biotechnology are biocompatible magnetic nanoparticles \cite{pankhurst_applications_2003}.}
These particles can be tailored to a particular application by engineering of their size, composition, and coating \cite{lu_magnetic_2007}.
Moreover, magnetic nanoparticles can be attracted by a magnet and externally visualized \cite{pankhurst_applications_2003}, which can help detection and supervision.
Applications of magnetic nanoparticles include tissue engineering \cite{durr_magnetic_2016}, biosensing \cite{giouroudi_microfluidic_2013}, imaging \cite{gleich_tomographic_2005}, remotely stimulating cells \cite{dobson_remote_2008}, waste-water treatment \cite{raj_coconut_2015}, and drug delivery \cite{tietze_efficient_2013}.

In the context of MC, the use of magnetic nanoparticles as information carriers has been considered in \cite{wicke_magnetic_2019,schafer_nd_2018,schafer_analytical_2019,schafer_transfer_2020} and \cite{kisseleff_magnetic_2017}, where the benefits of attracting them towards a receiver are theoretically evaluated and the design of a wearable device for detecting them is proposed, respectively.
Furthermore, \changed{based on the conference version of this work \cite{unterweger_experimental_2018}, the design and characterization of receiver \cite{bartunik_novel_2019,ahmed_characterization_2019,bartunik_comparative_2020} and transmitter devices \cite{bartunik_amplitude_2020,schlechtweg_magnetic_2019} for MC systems using magnetic nanoparticles have been investigated.
However, a comprehensive communication-theoretical analysis of the complete experimental magnetic nanoparticle based MC system has not been provided, yet.}

Similar to \cite{farsad_novel_2017,khaloopour_experimental_2019,tuccitto_reactive_2018}, in this paper, we present a testbed for in-vessel MC.
Our setup differs in that it uses specifically designed superparamagnetic iron oxide nanoparticles (SPIONs) as information carriers, which are biocompatible, clinically safe, and do not interfere with other chemical processes and thus might be attractive for applications such as monitoring of chemical reactors where particles stored in a reservoir could be released upon an event like the detection of a defect which is then communicated to a central control station for further processing.
In the proposed testbed, SPIONs are injected and transported along a propagation tube by fluid flow which is established by a peristaltic pump.
The propagation tube runs through the receiver where the magnetic susceptibility of the mixture of water and SPIONs within a section of the tube can be non-intrusively determined.
The magnetic susceptibility measured in the tube section is proportional to the concentration of the particles within the section.
This proportionality is more amenable to mathematical analysis compared to observing the pH in~\cite{farsad_novel_2017,jamali_diffusive_2018,grebenstein_biological_2019}, which non-linearly depends on the underlying proton concentration.

The contributions of this paper can be summarized as follows:
\begin{itemize}
    \item
    We present an experimental system for MC based on the flow-driven transport of SPIONs in a tube.
    All components of the system are described in detail which was not possible in \cite{unterweger_experimental_2018} due to space constraints.
    \item
    Extending \cite{unterweger_experimental_2018}, we provide a physical characterization of the system regarding the relevant effects for particle injection, propagation by flow in the tube, and reception by the susceptometer.
    In MC terms, we motivate the model of a transparent receiver and characterize the pulse shaping by the injection via an initial volume distribution \cite{noel_channel_2016}.
    \item
    Motivated by the physical characterization, we develop a \changed{novel} parametric model for the \changed{macroscopic} system's channel impulse response (CIR) providing insight into the flow-driven transport and the receiver's physical properties.
    We validate the applicability of our model by fitting its parameters to measurement data of the CIR showing a good agreement despite the simplifications needed for analytical tractability.
    \item
    To demonstrate successful information transmission, we show that reliable detection of on-off keying (OOK) is possible for transmission distances of up to \SI{40}{\centi\meter} and a symbol duration of \SI{1}{\second}.
    To this end, we propose and evaluate symbol sequence estimation by applying 1) an optimal detection rule assuming a linear pulse-amplitude modulation (PAM) model and 2) a model-agnostic heuristic detection rule based on the signal increases and decreases following symbol intervals with injections and idle times, respectively.
\end{itemize}

The rest of this paper is organized as follows.
In Section~\ref{sec:system_description}, we explain the testbed and its components as well as physical preliminaries.
In Section~\ref{sec:mathematical model}, we propose a \changed{closed-form} CIR model.
In Section~\ref{sec:demodulation}, we describe the employed signal processing and detection algorithms.
In Section~\ref{sec:experiments}, we present experimental data.
Finally, in Section~\ref{sec:conclusion} we conclude the paper and provide directions for future work.

\section{Magnetic Nanoparticle-Based Testbed}
\label{sec:system_description}

\begin{figure*}[!t]
    \centering
    \subfloat[]{\label{fig:photo}%
        \includegraphics[]{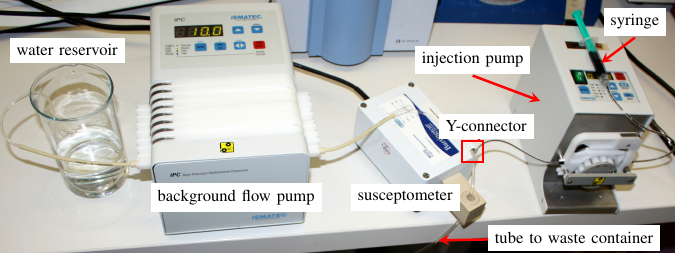}
    }\\
    \subfloat[]{\label{fig:injection}%
        \includegraphics[]{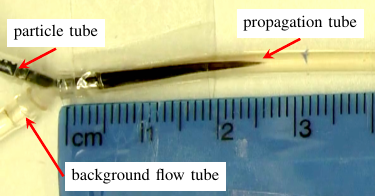}%
    }\hfil
    \subfloat[]{\label{fig:particle concept}%
        \includegraphics[width=8cm]{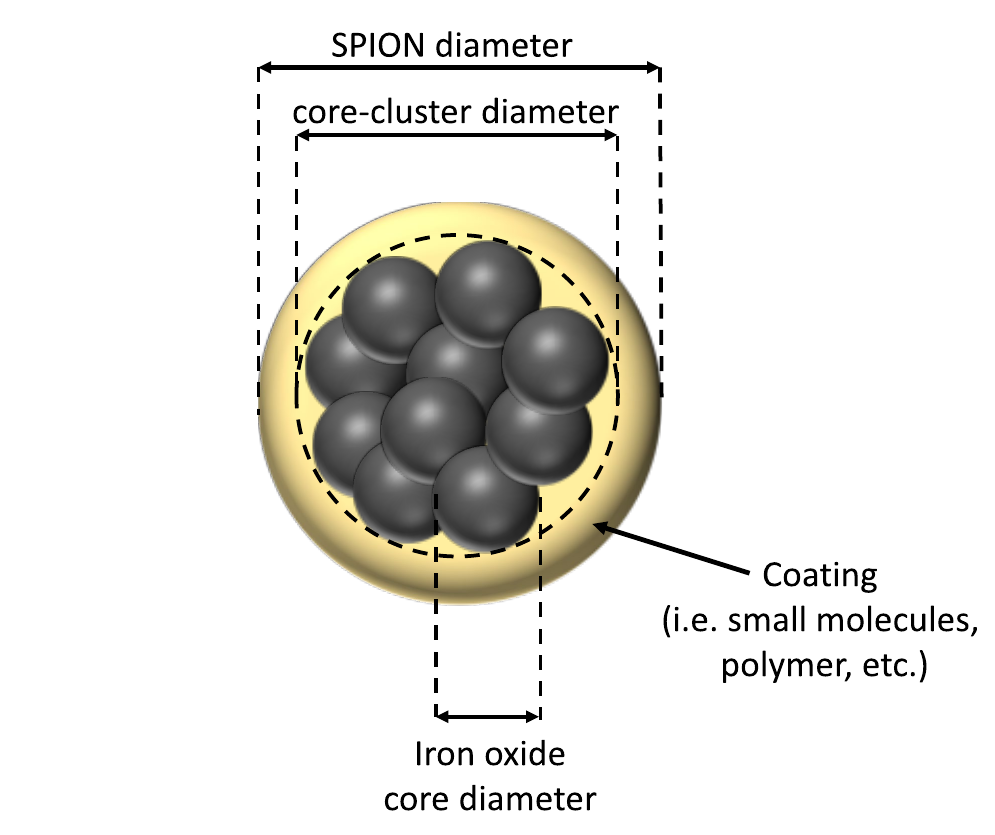}%
    }
    \caption{SPION testbed.
        (a) Photograph of the testbed showing the water reservoir, the background flow pump, the susceptometer, the pump used for injection, the syringe holding the suspension of SPIONs, and flexible plastic tubes connecting the components.
        The waste container below the table is not shown.
        (b) Photograph of the Y-connector with elongated SPION suspension right after injection for a slow background flow of $\Qb=\SI{1}{\milli\liter/\minute}$.
        The injected SPION suspension is elongated by the flow profile.
        (c) Schematic SPION composition consisting of iron oxide cores forming a core-cluster stabilized by a polymer.
    }
    \label{fig:aufbau}
\end{figure*}

\newcommand{\chiStock}{\SI{3e-3}{}}
\newcommand{\fittedRadius}{\SI{2.03}{\milli\meter}}
\newcommand{\fittedLength}{\SI{5.7}{\milli\meter}}
\newcommand{\nominalLength}{\SI{5}{\milli\meter}}
\newcommand{\RXinnerRadius}{\SI{5}{\milli\meter}}
\newcommand{\tuberadius}{\SI{0.75}{\milli\meter}}
\newcommand{\hydrad}{\SI{24.5}{\nano\meter}}
\newcommand{\particlemass}{\SI{2.5e-19}{\kilogram}}
\captionsetup{position=top}
\begin{table}
    \caption{System Parameters}
    \label{tab:system parameters}
    \centering
    \subfloat[Particle Properties]{
        \begin{tabular}{ll}
            \toprule
            Parameter                               & Value\\
            \midrule
            Hydrodynamic particle radius            & \hydrad\\
            Suspension iron stock concentration     & \SI{7.89}{\milli\gram/\milli\liter}\\
            Suspension magnetic susceptibility      & \chiStock{} (SI units) \\
            Particle mass                           & \particlemass{} \\
            \bottomrule
        \end{tabular}
    }
    
    \subfloat[Testbed Settings]{
        \begin{tabular}{ll}
            \toprule
            Parameter                               & Range/Value\\
            \midrule
            Tube radius particle injection          & \SI{0.40}{\milli\meter}	\\
            Tube radius background flow	$a$	        & \tuberadius{}	\\
            Flow rate particle injection $\Qp$      & \SI{5.26}{\milli\liter/\minute}	\\
            Flow rate background flow $\Qb$	        & \SI{5}{\milli\liter/\minute}	\\
            Volume particle injection $\Vi$		    & \SI{17.3}{\micro\liter}	\\
            Symbol duration $T$		        & \SI{1}{\second}	\\
            Propagation distance $d$                & \SIrange{5}{40}{\centi\meter}\\
            \bottomrule
        \end{tabular}
    }
\end{table}
\captionsetup{position=bottom}
%
\changed{In this section, we first describe the proposed overall testbed, and then we discuss each of its components individually.
A representative photograph of the entire system is shown in Fig.~\ref{fig:photo} and works as follows.
A tube with a constant background flow maintained by a background flow pump fed by a water reservoir serves as channel.
For signal transmission, SPIONs stored in a syringe are injected into the background flow by an injection pump which is connected to the main tube via a Y-connector.
The tube with the background flow then passes through a magnetic susceptometer for signal detection before emptying in a waste container.
The received signal is generated by the susceptometer detecting the SPIONs mixed in the background flow.
A close-up of the injection process is shown in Fig.~\ref{fig:injection}, and a sketch of the composition of one individual SPION is shown in Fig.~\ref{fig:particle concept}.
 The system parameters are summarized in Table~\ref{tab:system parameters} and further described in the following.
 }

\subsection{Testbed Components}
In this subsection, we describe the physical setup of the testbed components, including the information-carrying particles, the transmitter, the propagation channel, and the receiver.

\subsubsection{Carrier}
\label{sec:carrier}
We use a specific type of magnetic nanoparticles as information carriers which are referred to as SPIONs.
\changed{SPIONs consist of clusters of iron oxide particles and a coating for chemical stability, see Fig.~\ref{fig:particle concept}.
For producing clusters of iron oxide particles, there is a multitude of possibilities, including thermal, sol-gel, electrochemical, and precipitation approaches.
One of the fastest, simplest, and most efficient methods to synthesize SPIONs is the coprecipitation technique in alkaline media as it was first proposed by the authors of~\cite{massart_preparation_1981,khalafalla_preparation_1980} in the early 1980s.}
The first and most crucial step of this synthesis is the precipitation of magnetite ($\mathrm{Fe}_3\mathrm{O}_4$) from ferric ($\mathrm{Fe}^{3+}$) and ferrous ($\mathrm{Fe}^{2+}$) salts with a stoichiometric ratio of 2:1 ($\mathrm{Fe}^{3+}/\mathrm{Fe}^{2+}$) in an inert atmosphere at a basic pH:
\begin{equation*}
2\mathrm{Fe}^{3+} + \mathrm{Fe}^{2+} + 8\mathrm{OH}^- \rightarrow \mathrm{Fe}_3\mathrm{O}_4 + 4\mathrm{H}_2\mathrm{O},
\end{equation*}
consuming hydroxide ($\mathrm{OH}^-$) with water ($\mathrm{H}_2\mathrm{O}$) as byproduct.

For our synthesis, we use ammonia to start the formation of the particles.
Since nanoparticles in general and our SPIONs in particular are featured with a small size, they possess a large surface to volume ratio.
The resulting high surface energy renders the particles thermodynamically unstable and is responsible for their tendency to minimize their energy by agglomeration.
\changed{In order to avoid this behavior, a suitable stabilization mechanism is required.
To this end, the clusters of iron oxide particles are coated.}
Generally, stabilization can be achieved by small molecules, polymers, and proteins.
It is common to all of them to produce repulsion either by electrostatic, by steric, or by electrosteric means.
Most colloidal dispersions possess an electric surface charge which, depending on the material and the dispersion medium, gives rise to electrostatic stabilization.
However, certain tradeoffs apply for the coating of SPIONs in MC.
First, the synthesis and coating has to be designed to make the particles as large as possible, in order to be able to generate a large signal for detection.
However, the larger the particles, the more they are prone to sedimentation.
In addition the coating material should provide the particles not only with stability against agglomeration but also render them inert against the components of the testbed.
For these reasons, we used SPIONs with lauric acid as a stabilizing agent~\cite{zaloga_development_2014}. 

\changed{
The SPIONs are dispersed in an aqueous suspension and stored in a syringe, which is connected to a tube with an inner radius of \SI{0.4}{\milli\meter}.
Their properties are as follows, see also Table~\ref{tab:system parameters}.
The SPIONs have a hydrodynamic radius of \hydrad{} (measured by dynamic light scattering), an iron stock concentration of \SI{7.89}{\milli\gram/\milli\liter} (measured by atomic emission spectroscopy), a susceptibility of \chiStock{} (dimensionless in SI units, measured with the susceptometer of the testbed), and an estimated individual particle mass of approximately \particlemass{}.
The estimated concentration of SPIONs in aqueous suspension is
approximately \SI{4d13}{particles/\milli\liter} with a viscosity of \SI{1.34}{\milli\pascal\second} close to that of water which is at \SI{1.08}{\milli\pascal\second} (measured with a viscometer at room temperature).
}

\subsubsection{Transmitter}
\label{sec:system_description_transmitter}
The movement of the particles through the tube is established with a computer controlled peristaltic pump (Ismatec Reglo Digital, Germany), which can provide discrete pumping actions at a flow rate of \SI{5.26}{\milli\liter/\minute} (maximal speed), injecting a dosage volume of $\Vi=\SI{17.3}{\micro\liter}$ of SPION suspension.
\changed{We note that intersymbol interference (ISI) can be reduced by minimizing the injection duration at the expense of signal strength either by increasing the injection speed or decreasing the injection volume.
Hence, the injection speed was set to its maximal value to reduce ISI and the injection volume has been adjusted manually so as to achieve a strong signal for the following analysis}.

The end of the tube with the particles is joined via a Y-connector with another tube of radius \SI{0.75}{\milli\meter} providing a background flow, see Fig.~\ref{fig:injection}.
The constant background flow of water has a flow rate of \SI{5}{\milli\liter/\minute} and is maintained by a second pump (Ismatec IPC, Germany).

Discrete pumping actions with a minimum adjustable separation of \SI{1}{\second} are realized by a custom LabVIEW (National Instruments, Austin, Texas, USA) graphical user interface (GUI) that triggers the particle pump.

\subsubsection{Propagation Channel}
\label{sec:system_description_channel}
The flow rate in the tube channel is the sum of the rates of the background flow and the particle injection.
It is hence time-dependent and amounts to \SI{10.26}{\milli\liter/\minute} during particle injection and \SI{5}{\milli\liter/\minute} in the remaining time.
When particles are pumped into the channel by the transmitter, then the resulting particle cloud is entrained by the flow and simultaneously diluted and elongated, see Fig.~\ref{fig:injection}.

The length of the propagation channel is variable but was set to \SIlist{5;10;20;40}{\centi\meter} for the results shown in Section~\ref{sec:experiments}.

\subsubsection{Receiver}
\label{sec:receiver}
At the end of the propagation channel, the tube runs through the air core of an MS2G Bartington susceptometer coil (inner diameter: \SI{10}{\milli\meter}, length: \SI{20}{\milli\meter}, specified sensitive length: \SI{5}{\milli\meter}).
When the magnetic particles are within the detection range of the susceptometer, an electrical signal $\chi(t)$ is generated.
This signal is proportional to the number of SPIONs that are within the detection range at a specific time instance.
After the particles have passed through the receiver, they are collected in a waste bin together with the water from the background flow.
Water has a small negative magnetic susceptibility of about \SI{-9.0e-6}{} (SI units) \cite[Appendix~E]{coey_magnetism_2010}.
Hence, the magnitude of the magnetic susceptibility of water is much smaller than that of the considered SPION suspension $\chiref=\chiStock{}$ (SI units).

The susceptibility changes measured at the receiver were recorded by use of the software Bartsoft 4.2.1.1 (Bartington Instruments, Witney, UK) provided by the manufacturer of the susceptometer.
The susceptometer is a reliable and convenient commercial device for characterizing the magnetic susceptibility of SPION suspensions in the laboratory.
Nevertheless, we note that our current use case of measuring time signals with a short sampling period in the order of \SI{0.1}{\second} is outside of the usual mode of operation of this device which has been designed for one-time measurements of bulk probes under stationary conditions, see \cite{bartunik_comparative_2020} for the evaluation of a custom susceptometer design.
Hence, care has to be taken when interpreting the measured signal as magnetic susceptibility since we operate the susceptometer outside of its specification by evaluating the output signal for a spatially varying SPION concentration over time.

\subsection{General Considerations}
In this subsection, we provide some background on the relevant physical effects affecting the measurement signals.
These considerations will guide our mathematical modeling efforts in Section~\ref{sec:mathematical model}.

\subsubsection{Turbulent or Laminar Flow}
Fluid flow can be categorized as either laminar or turbulent.
This categorization determines the appropriate mathematical model to be used.
While laminar flow is prevalent in microfluidic applications, turbulent flow is encountered in macroscale ducts in the size range of several centimeter.
The relevant parameter which predicts a transition from laminar to turbulent flow is the Reynolds number $\Reyn\in\reals^+$ which is defined as \cite[Chapter~3]{deen_introduction_2016}
\begin{equation}
\label{eq:reynolds number}
\Reyn = \frac{a \cdot \vmax}{\visc},
\end{equation}
where $a$ is the tube radius, $\vmax$ is the maximum flow velocity in the tube and can be computed as $\vmax=2\veff$ with the area-averaged velocity in the tube $\veff=\Qb/(\pi a^2)$, $\visc$ is the kinematic viscosity of the liquid, and $\Qb$ is the background flow rate, i.e., the total flow rate after injection.
For a circular duct, a value of $\Reyn\approx\num{2100}$ is critical for the transition from laminar to turbulent flow, see \cite[Chapter~3]{deen_introduction_2016}.
For the testbed parameters in Table~\ref{tab:system parameters}, we find $\veff=\SI{47.2}{\milli\meter/\second}$.
Thus, using the kinematic viscosity of water $\visc=\SI{e-6}{\meter^2/\second}$ \cite[Chapter~1]{deen_introduction_2016}, we obtain $\Reyn=\SI{70.7}{}<\num{2100}$ and hence expect fully laminar flow.
By the reasoning above, we would expect a transition to turbulent flow for an increase of the Reynolds number in \eqref{eq:reynolds number} by a factor of \num{30}.
For example, all other parameters held equal, we would expect a transition to turbulent flow at an effective flow speed of $\veff=\SI{1410}{\milli\meter\per\second}$ ($\Qb=\SI{150}{\milli\liter\per\minute}$) or for a tube radius of \SI{22.5}{\milli\meter}.
We note that additional turbulence could also be caused by obstacles in the tube.
\changed{Furthermore, turbulences might potentially occur during injection.
However, these turbulent effects can be assumed to be spatially limited close to the injection site and temporally limited to the duration of injection.
Hence, we still expect laminar flow to dominate the SPION transport to the receiver.}

\subsubsection{Diffusion}
\label{sec:diffusion}
In general, the particle motion is governed by both diffusion and transport by fluid flow, assuming a fully dilute aqueous SPION suspension.
The relative importance of diffusion compared to the transport by fluid flow over a distance of $d$ can be quantified by a dispersion factor $\dispfac\in\reals^+$ which is defined as \cite[Section~II.B]{jamali_channel_2019}
%
\begin{equation}
\dispfac = \frac{d D}{\chardist^2 \vel}
\end{equation}
where $D$ is the \changed{molecular} diffusion coefficient of the SPIONs, $\chardist$ is a characteristic distance over which the flow velocity varies, here chosen as $\chardist=a/10$, and $\vel$ is an effective velocity, here chosen as $\vel=\veff$.
When $\dispfac\ll 1$ and $\dispfac\gg 1$, over a distance of $d$, flow and diffusion dominate the transport, respectively.

The \changed{molecular} diffusion coefficient can be estimated as \cite[Section~II]{jamali_channel_2019}
\begin{equation}
\label{eq:diffusion coefficient}
D = \frac{\kB\Tm}{\zeta},
\end{equation}
where $\kB\Tm=\SI{4.11e-21}{\joule}$ is a characteristic energy with Boltzmann constant $\kB$ and the temperature of the liquid medium $\Tm=\SI{298}{\kelvin}$, and $\zeta=6\pi\eta\Rp=\SI{5.18e-10}{\kilogram\per\second}$ is the friction coefficient.
Here, $\eta=\SI{1e-3}{\pascal\second}$ is the dynamic viscosity of the liquid medium (water) and $\Rp=\SI{24.5}{\nano\meter}$ is the SPION radius.
From \eqref{eq:diffusion coefficient}, we estimate \changed{$D=\SI{8e-12}{\meter^2/\second}$} for the considered SPIONs.
Hence, for the testbed parameters in Table~\ref{tab:system parameters}, we obtain \changed{$\dispfac=\num{1.2e-2}$} (for $d=\SI{40}{\centi\meter}$).
This value is \changed{two} orders of magnitude smaller than $1$ and therefore the impact of diffusion is assumed to be negligible over the considered distances $d\leq\SI{40}{\centi\meter}$.
By the reasoning above and taking $\dispfac=1$ as critical value, we would expect diffusion to have a noticeable impact for an increase of the dispersion factor by a factor of \num{84}.
All other parameters held equal, this would be the case for a decrease of the effective flow velocity to $\veff=\SI{0.56}{\milli\meter\per\second}$ ($\Qb=\SI{0.06}{\milli\liter\per\minute}$), an increase of the distance to $d=\SI{30}{\meter}$, a smaller tube radius of $a=\SI{0.082}{\milli\meter}$, or a much larger diffusion coefficient of $D=\SI{8e-10}{\meter^2\per\second}$.
We note that the diffusion coefficient could effectively increase by particle-particle interactions or turbulence~\cite{deen_introduction_2016}.

\subsubsection{Injection}
\changed{The injected SPION suspension is in aqueous phase with a viscosity similar to that of water, especially when mixed with the background flow.
Hence, after injection, a one-phase flow is expected, rather than a two-phase flow, as would be the case for, e.g., an oily suspension in water.}
During injection, we have two joining flows.
For an injection flow rate of $\Qp=\SI{5.26}{\milli\liter\per\minute}$ and a background flow rate of $\Qb=\SI{5}{\milli\liter\per\minute}$, we have a net flow rate of $\Qb+\Qp=\SI{10.26}{\milli\liter\per\minute}$ during injection and a flow rate of $\Qb=\SI{5}{\milli\liter\per\minute}$ when not injecting.
This corresponds to variations of the effective flow velocity between $\SI{47.2}{\milli\meter\per\second}$ and $\SI{96.8}{\milli\meter\per\second}$.
Following an injection, the increased flow velocity occurs for the injection duration of \SI{197}{\milli\second} and in principle affects the signal generated by all previous injections.
By considering the flow rates, we can estimate the injection depth from the top to the bottom of the pipe by $2a\cdot\Qp/(\Qp+\Qb)$ \cite{deen_introduction_2016}, e.g., for $\Qp=\Qb$ we would have an injection depth of half the pipe diameter and for $\Qp\ll\Qb$ the injection depth would be close to 0.
For the testbed parameters given in Table~\ref{tab:system parameters}, we can determine the injection depth as $\SI{0.77}{\milli\meter}$, i.e., we expect the injected SPION suspension to reside in the upper half of the tube of diameter $2\cdot\tuberadius{}$.
In fact, the particle volume distribution after injection determines the received signal as the SPION transport is expected to be deterministic driven by the flow.

\subsubsection{Gravity}
Another \changed{potentially relevant effect for the transport of larger particles in a macroscopic system, especially for larger distances, is gravity.}
The gravitational force on an individual SPION can be determined as $F=mg=\SI{2.5e-18}{\newton}$ where $m=\SI{2.5e-19}{\kilogram}$ is the particle mass, see Table~\ref{tab:system parameters} and $g=\SI{9.81}{\meter\per\second^2}$ is the gravitational constant \cite{deen_introduction_2016}.
The resulting drift velocity due to gravity can then be determined as $\vel=F/\zeta=\SI{4.7}{\nano\meter\per\second}$ \cite{jamali_channel_2019} where $\zeta$ is the friction coefficient in Section~\ref{sec:diffusion}.
When we consider a vertical displacement by gravity of $a/10=\SI{75}{\micro\meter}$ to be relevantly large, then we obtain a time duration after injection of $a/10/\vel=\SI{4.4}{\hour}$ where gravity begins to matter.
Since the considered CIRs have a duration of less than \SI{1}{\minute}, we expect that the effect of gravity on an individual SPION is negligible.
On the other hand, considering a time duration of \SI{1}{\minute} as critical, gravity would begin to matter for a particle mass of \SI{6.61e-17}{\kilogram} which would correspond to a radius larger than \SI{176}{\nano\meter} assuming the same mass density, i.e., assuming the SPION mass is proportional to $\Rp^3$.
\changed{%
We note that the effective particle size could increase by particle-particle interaction.
In cases where gravity matters, the particle transport by the fluid flow has to be considered jointly with the downwards motion resulting from gravity, see also \cite{schafer_analytical_2019,schafer_nd_2018}.
}

\subsubsection{Magnetic Susceptibility}
\label{sec:magnetic susceptibility}
Finally, we consider the receiver device and the measured received signal.
The magnetic susceptometer used as receiver is a device comprising an electromagnetic coil with an air-core used as measuring space in an electric resonance circuit.
The susceptometer is designed for measuring the magnetic susceptibility of bulk material probes large enough to fill the coil (bulk magnetic susceptibility).
This bulk magnetic susceptibility is proportional to the change of inductance resulting from inserting the bulk material into the coil which can be measured, e.g., by examining the resonance frequency of the coil~\cite{bartunik_comparative_2020}.
For example, the SPION suspension used in this testbed has a bulk susceptibility of $\chiref$ and a bulk mixture of the SPION suspension with water at a ratio of $c\in[0,1]$ can be expected to yield a susceptibility of $\chim=\chiref\cdot c$.
However, for locally varying concentrations as is the case for our testbed where the SPION suspension is being transported and dispersed in the water by the fluid flow, the susceptometer output does only reflect an average susceptibility.

Motivated by the above analysis, in the following section, a mathematical model is established accounting for the transport by laminar flow where the received signal is characterized by the injection and a spatially weighted integral of the local susceptibility.

\section{Mathematical Signal Model}
\label{sec:mathematical model}
\begin{figure}[!t]
    \centering
    \includegraphics{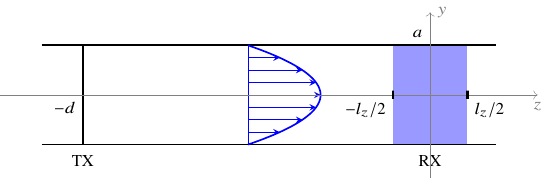}
    \caption{
        Sketch of the simplified system model consisting of a straight tube with radius $a$, shown for a $(z,y)$-cut with transmitter node TX concentrated at axial coordinate $z=-d$ and transparent receiver node RX weighting the SPION concentration.
        The laminar profile in \eqref{eq:flow profile} is schematically shown by velocity vectors.
    }
    \label{fig:abstract system}
\end{figure}

%
In this section, first the modeling assumptions for the channel, transmitter, and receiver are described and then the CIR, i.e., the expected received signal for one single injection, is derived.
\changed{We note that the CIR, normalized with respect to the injected number of particles, can also be interpreted as the probability of observing an individual signaling particle at a given time.}
For quick reference, a sketch of the assumed abstract system model is shown in Fig.~\ref{fig:abstract system}.

\subsection{Modeling Assumptions}
First, we will describe the flow-driven propagation in the tube channel, then we characterize transmitter and receiver.
In the following, we will use cylindrical coordinates for position $\bm{x}=(\rho,\phi,z)$, where $\rho$, $\phi$, and $z$ are the axial distance, azimuth, and axial coordinates, respectively.

\subsubsection{Channel}
In general, the flow at the injection site is complicated and time-variant as alluded to in Section~\ref{sec:system_description_channel}.
Moreover, in our testbed, the tube is not perfectly straight and is not infinite but inherently also includes the Y-connector used for injection.
This leads to a complicated flow profile in general even without injections.
Nevertheless, to simplify the analysis and as the deviations over distances on the order of the inner tube diameter are small, we will assume laminar flow in a straight tube of circular cross-section as the main particle transport mechanism \changed{after injection}.
In this case, the non-uniform flow velocity profile is well known to be \cite{deen_introduction_2016}
%
\begin{equation}
\label{eq:flow profile}
\vel(\rho) = \vmax \cdot \left( 1 - \frac{\rho^2}{a^2} \right),
\end{equation}
where $\rho$ is the radial distance from the central axis of the tube.

Then, the concentration develops over time and space based on the following model for flow-driven transport~\cite{jamali_channel_2019}
\begin{equation}
\label{eq:c flow}
c(\bm{x},t) = \cin(\bm{x}-\vel(\rho)t\cdot \bm{e}_z),
\end{equation}
where $\cin(\bm{x})$ is the assumed initial spatial particle distribution and $\bm{e}_z$ is the unit vector in $z$ direction.
The concentration satisfies $\iiint c(\bm{x},t) \de V = \Vi, \forall t$ where $V$ is the volume of the infinite cylinder.

\subsubsection{Transmitter}
\label{sec:tx spec}
The transmitter is physically realized by the injection pump and the Y-shaped tube connector, see Section~\ref{sec:system_description_transmitter}.
However, for modeling, we will focus on the initial particle distribution within an infinite cylinder which is created by the injection process, i.e., the volume distribution of SPIONs right after injection.

As first order approximation, we will assume that the initial distribution can be modeled as being axially concentrated at the site of injection\footnote{%
We note that a more accurate model could be obtained, for example, by numerical simulation of the injection process and evaluating the obtained initial volume distribution \cite{drees_efficient_2020}.
    However, as this numerical simulation \changed{is computationally demanding and} does not directly give theoretical insight, in this paper, we focus on a simple parametric model for the initial distribution. \changed{Nevertheless, studying the injection process and the resulting pulse shaping options in more detail constitutes an interesting topic for future work.}%
}.
With this assumption, the initial concentration in space can be mathematically expressed as
\begin{equation}
\label{eq:cin}
\cin(\bm{x}) = \Vi \cdot f(\rho,\phi) \cdot \delta(z+d),
\end{equation}
where $f(\rho,\phi)$ is the distribution in the cross-section of the tube  and $\delta(z)$ is the Dirac delta function.
The distribution in the cross-section is normalized to $\iint f(\rho,\phi)\de A=1$, where $A$ denotes the area of the tube cross-section.
The transmitter is at $z=-d$, see Fig.~\ref{fig:abstract system}, and the injection volume is assumed to be concentrated at position $z=-d$.
We note that in general the initial distribution in \eqref{eq:cin} depends on the angular coordinate $\phi$.
However, the received signal will not depend on the distribution over $\phi$ because of the geometrically symmetric arrangement of the receiver surrounding the tube.
Therefore, in the following, we will introduce some definitions to describe the particle distribution over the radial coordinate.

\paragraph{Definitions}
The radial distribution is given by
\begin{equation}
\label{eq:frho def}
f_\rho(\rho) = \int_0^{2\pi} f(\rho,\phi)\rho\de\phi
\end{equation}
and its cumulative distribution function is given by $F_\rho(\rho) = \int_{-\infty}^\rho f_\rho(\tilde{\rho})\de\tilde{\rho}$.
For convenience, we also define an auxiliary distribution as
\begin{equation}
\label{eq:fs spec}
f_s(s) = \frac{a}{2\sqrt{s}} \cdot f_\rho(a\sqrt{s})
\end{equation}
which satisfies $\int_0^1 f_s(s) \de s=1$ and where $s=\rho^2/a^2$.
The corresponding cumulative distribution function satisfies $F_s(s) = F_\rho(a\sqrt{s})$ according to \eqref{eq:fs spec}.

\paragraph{Special Cases}
Let's consider two common models for injection as important special cases \cite{levenspiel_chemical_1999}.
First, for a uniform particle distribution, $f(\rho,\phi)=1/(\pi a^2)$, we obtain $f_\rho(\rho)=2\rho/a^2, 0\leq\rho\leq a$ and $f_s(s) = 1, 0\leq s\leq 1$.
Second, for a particle distribution proportional to the flow profile in \eqref{eq:flow profile}, $f(\rho,\phi)=\frac{2}{\pi a^2} (1-\frac{\rho^2}{a^2})$, we obtain $f_\rho(\rho) = \frac{4}{a^2}\rho\cdot(1-\frac{\rho^2}{a^2})$ and $f_s(s)=2(1-s)$.

\paragraph{Parametric Initial Distribution}
To generalize from these two important special cases, we propose the following distribution in $s$
\begin{equation}
    \label{eq:initial distribution}
    f_s(s) = \frac{1}{B(\alpha,\beta)}\cdot s^{\alpha-1} \cdot (1-s)^{\beta-1}
\end{equation}
which is the Beta distribution \cite{papoulis_probability_2002} with $0\leq s\leq 1$.
The parameters shaping the Beta distribution are $\alpha\geq1$ and $\beta\geq1$ and the normalization is given by $B(\alpha,\beta) = \frac{\Gamma(\alpha)\Gamma(\beta)}{\Gamma(\alpha+\beta)}$, where $\Gamma(\cdot)$ is the Gamma function.

The Beta distribution is limited to the range $[0,1]$ which makes it suitable for modeling the range of parameter $s$.
We can also recover the uniform distribution for $(\alpha=1,\beta=1)$ and the distribution proportional to the flow profile for $(\alpha=1,\beta=2)$, where $B(1,\beta)=1/\beta$.
Moreover, for $\alpha\geq1$ and $\beta\geq1$, $f_s(s)$ is unimodal with adjustable peak position and peak width as is needed for our testbed where the SPION mass is concentrated around a certain radial position due to the injection.
These properties make the Beta distribution a good candidate for modeling the initial distribution following injection for this testbed.

For future reference, by using \eqref{eq:fs spec}, for $f_s(s)$ in \eqref{eq:initial distribution}, we can write the corresponding radial distribution as
\begin{equation}
    \label{eq:radial distribution}
    f_\rho(\rho) = \frac{2}{a B(\alpha,\beta)} \cdot \left(\frac{\rho}{a}\right)^{2\alpha-1} \cdot \left(1-\frac{\rho^2}{a^2}\right)^{\beta-1}.
\end{equation}

%



\subsubsection{Receiver}
The physical detection device is given by the susceptometer.
Motivated by the physical reception mechanism described in Section~\ref{sec:magnetic susceptibility}, we assume the following received signal model
\begin{equation}
\label{eq:first susceptibility model}
\chi(t) = \chiref \cdot \iiint_{\mathbb{R}^3} w(\bm{x})\cdot c(\bm{x},t)\de V
\end{equation}
where $w(\bm{x})$ is a weighting function which can be interpreted as a receiver window and is normalized as $\iiint_{\mathbb{R}^3} w(\bm{x})\de V=1$.
	For example, for $c(\bm{x},t) = 1, \forall \bm{x}, t$ we have $\chi(t)=\chiref$.
The receiver mechanism underlying \eqref{eq:first susceptibility model} can be understood as a transparent receiver~\cite{noel_channel_2016}.

In the following, for simplicity, we will assume that the receiver weighting function is a rectangular window only dependent on $z$, leaving a three-dimensional characterization of the weighting function for future work.
Hence, for our modeling, the rectangular receiver weighting function is given as
\begin{equation}
\label{eq:rectangular window}
w(\bm{x}) = \frac{1}{\pi a^2 l_z} \cdot\rect\left(\frac{z}{l_z}\right)
\end{equation}
where $l_z$ is the window length and $\rect(z)=1$ for $|z|\leq 0.5$, and $\rect(z)=0$ otherwise, i.e., the receiver is centered at axial position $z=0$, see Fig.~\ref{fig:abstract system}.

This concludes our list of modeling assumptions.
As will be shown, with these assumptions, we can capture the main characteristics of the measurement signals.
For example, when due to the injection more particles are concentrated close to the central axis than close to the boundary of the tube, a faster decay of the received signal over time and a larger peak can be expected due to the laminar flow profile in \eqref{eq:flow profile}.
The general CIR behavior under the given modeling assumptions is investigated in the following.

\subsection{Channel Impulse Response}
\label{sec:CIR equation}
In this subsection, we use the modeling assumptions summarized in the previous subsection to derive a compact closed-form CIR expression usable for fitting measurement data.
To this end, we assume the abstract system model shown in Fig.~\ref{fig:abstract system}, where the injection is simplified to a release concentrated at $z=-d$ and the receiver applies a spatially-weighted integral of the particle concentration resulting from the flow-driven transport similar to \cite{wicke_modeling_2018}.

\subsubsection{General Case}
Using \eqref{eq:c flow} and \eqref{eq:cin} in \eqref{eq:first susceptibility model}, as shown in the Appendix, we can express the CIR as follows
\begin{equation}
\label{eq:spawc model}
    h(t) = \myscal \cdot \frac{d}{l_z}\cdot\left(F_s\left(1 - \frac{d-l_z/2}{\vmax t}\right) - F_s\left(1 - \frac{d+l_z/2}{\vmax t}\right)\right),
\end{equation}
where $\myscal=\chiref \frac{\Vi}{\pi a^2 d}$ is a distance-dependent dimensionless scaling factor.

We note that with $f_s(s)$ in \eqref{eq:initial distribution}, \eqref{eq:spawc model} is a generalization of the solution provided in \cite{unterweger_experimental_2018}.
In particular, \eqref{eq:spawc model} reduces to \cite[Eq.~(5)]{unterweger_experimental_2018} for $\alpha=1$, $\beta\leftarrow\beta+1$ and $l_z=c_z$.
A numerical evaluation of \eqref{eq:spawc model} for different initial distributions is presented in Section~\ref{sec:cir model evaluation}.
\changed{We note that this numerical evaluation can be performed efficiently thanks to the closed-form expression of the CIR, which only involves the cumulative distribution function of the beta distribution.}

\subsubsection{CIR Analysis}
\label{sec:CIR modeling}
For the following analysis, for simplicity, we consider the case where $d\gg l_z$, i.e., the case when the transmitter-receiver distance is much larger than the receiver width.
Mathematically, we can employ the limit $l_z/d\to 0$ in \eqref{eq:spawc model}.
Then, we obtain
\begin{equation}
\label{eq:main equation}
h(t) = \myscal \cdot\frac{d}{\vmax t} \cdot f_s\left(1 - \frac{d}{\vmax t}\right),
\end{equation}
with $f_s(s)$ in \eqref{eq:initial distribution}.

Since \eqref{eq:main equation} provides a \changed{convenient} closed-form solution for the CIR independent of the receiver length, we will use it as our modeling function for fitting measurement results in Section~\ref{sec:experiments}.

For convenience, we plug \eqref{eq:initial distribution} into \eqref{eq:main equation} and arrive at
\begin{equation}
\label{eq:full model}
h(t) = \myscal\cdot\frac{1}{B(\alpha,\beta)}\cdot \left(1 - \frac{d}{\vmax t}\right)^{\alpha-1} \cdot \left(\frac{d}{\vmax t}\right)^{\beta}
\end{equation}
for $t\geq d/\vmax$ and $h(t)=0$ otherwise.
Interestingly, for large $t$, $h(t)$ decays as $1/t^\beta$.
This is related to the particle fraction at $\rho\to a$ which according to \eqref{eq:radial distribution} depends on $\beta$ but not on $\alpha$.

The initial delay of the received signal,
\begin{equation}
    \label{eq:tstart}
    t_0=\frac{d}{\vmax},
\end{equation}
predicted by \eqref{eq:full model}, is expected, as this is the time needed for particles initially placed in the center of the tube to travel distance $d$, i.e., a measure for the theoretical time offset until which no signaling particle has reached the receiver yet.
Moreover, we note from \eqref{eq:full model} that $h(t_0) = 0$ for $\alpha\neq1$ and $h(t_0) > 0$ for $\alpha=1$, i.e., there is a jump at $t=t_0$.

Finally, let us consider the position and the height of the peak (maximum) of the derived CIR.
For $\alpha>1$, the position of the peak of the CIR in \eqref{eq:full model} can be found at
\begin{equation}
\label{eq:tpeak}
\tpeak = \tstart\cdot\left(1 + \frac{\alpha-1}{\beta}\right),
\end{equation}
by equating the time derivative of $h(t)$ in \eqref{eq:full model} with zero and solving for $t$.
Interestingly, it can be observed via \eqref{eq:tstart} that for any $\alpha$ and $\beta$, $\tpeak$ is proportional to $d$.
By substituting \eqref{eq:tpeak} in \eqref{eq:full model}, the height of the peak follows as
\begin{equation}
\label{eq:hpeak}
\hpeak = \myscal\cdot\frac{1}{B(\alpha,\beta)} \cdot \frac{(\alpha-1)^{\alpha-1} \beta^\beta}{(\alpha+\beta-1)^{\alpha+\beta-1}},
\end{equation}
which is inversely proportional to distance $d$ for any combination of $\alpha$ and $\beta$.

As a summary of the CIR analysis, we conclude that for any choice of $\alpha$ and $\beta$, the peak height $\hpeak$ decays proportional to $1/d$ and $h(t)$, for large $t$, decays as $1/t^\beta$.

\subsubsection{Special Cases}
Let us consider again, the two special cases of a uniform particle distribution and that of a particle distribution proportional to the flow profile described earlier.
For these two cases, we expect a decay over time proportional to $1/t$ and $1/t^2$, respectively, see \cite[Chapter~15]{levenspiel_chemical_1999}.
This behavior is indeed recovered for $(\alpha=1,\beta=1)$ and $(\alpha=1,\beta=2)$ in \eqref{eq:full model}.

\section{Modulation, Channel Estimation, and Detection}
\label{sec:demodulation}
In this section, we discuss the communication and signal processing aspects of our testbed, i.e., the preprocessing of the raw data, channel estimation, and detection algorithms.

\subsection{Data Preprocessing}
Preprocessing of the data provided by the susceptometer is needed for a consistent postprocessing such as comparing with the developed CIR model and for detection.
By manual examination, it turns out that the employed susceptometer delivers samples at sampling times which are not perfectly regular and the absolute time is not synchronized with the injection.
For consistency, we employ a resampling by linear interpolation to 10 samples per second which corresponds to the average sampling times of the measurement data as provided by the susceptometer.

In the following, we denote the preprocessed susceptibility signal by $\chi[i]=\chi(i\dt)$ where $\dt=\SI{0.1}{\second}$ is the sampling interval and $i=0,1,2,\dots$.
Thereby, $\chi(t)$ is the underlying but inaccessible preprocessed time-continuous signal.

For time synchronization, we look for the start of the first occurrence of 10 consecutive samples surpassing a threshold chosen as one hundredth of the maximal observed signal amplitude in that measurement.
This time index is labeled as $\istart$ and the received signal $r[i]=\chi[i-\istart]$ is then used for further processing.
For future reference, the vector of received signal values is denoted by $\bm{r}$.

\subsection{Channel Estimation}
\label{sec:channel estimation}
The information to be detected is represented as follows.
We assume that information is represented by a sequence $\bm{a}\in\{0,1\}^{K}$ of OOK symbols with $a[k]\in\{0,1\}, k\in\{0,1,\dots,K-1\}$, where $K$ is the number of transmitted symbols.
This series of amplitude coefficients is then modulated on pumping pulses as described in Section~\ref{sec:system_description_transmitter}.
For PAM detection, we assume the following basic pulse amplitude modulation model for the noise-free received signal
\begin{equation}
    \label{eq:PAM model discrete}
    s[i;\bm{a},\bm{h}] = \sum_{k=0}^{K-1} a[k]\cdot h[i-kI]
\end{equation}
where $i$ is the sampling index, and $I=T/\dt=10$ is the oversampling factor corresponding to symbol interval $T$.
For convenience, the vector of noise-free received signal values is denoted by $\bm{s}(\bm{a},\bm{h})$.
In particular, $\bm{h}\in\reals^{N I}$ with entries $h[i], i=0,1,\dots,N I - 1$, are the samples of the CIR which can be obtained by estimation using training data, as described in the following, and $N$ is the memory length measured in numbers of symbols.

For sequence estimation, we need to know the overall CIR $h[i]$.
For our numerical results, we obtain this CIR by estimation based on training data sent at the start of transmission.
To this end, we denote the sequence of training symbols as $\atrain\in\{0,1\}^{\Ktrain}$ and the training samples of the received signal as $\rtrain\in\mathbb{R}^{\Ktrain I}$, where $\Ktrain$ is the length of the training sequence.
In a similar manner, $\bm{s}(\atrain,\bm{h})$ denotes the vector of model transmit training signal samples.
In the following, we describe two channel estimation schemes, one based on the CIR model developed in Section~\ref{sec:CIR equation} and another one which directly estimates all samples of the CIR.

\subsubsection{Model-based CIR Estimation}
For estimating the model parameters, we consider the following optimization problem
\begin{equation}
    \label{eq:nonlinear estimation}
    \hat{\alpha}, \hat{\beta}, \hat{\gamma} = \argmin_{\alpha,\beta, \scal>0} \, \lVert \rtrain - \bm{s}(\atrain, \scal\cdot\bm{h}(\alpha, \beta))\rVert^2,
\end{equation}
where the samples of the CIR are given as $\bm{h}(\alpha, \beta)=h(i\Delta t+t_0; \alpha,\beta), i=0,1,\dots,NI-1$.
This is a non-linear least-squares optimization problem with three parameters (scaling parameter $\scal$ and model parameters $\alpha$ and $\beta$).
The estimated CIR is then given by $\hat{\bm{h}}=\hat{\scal}\cdot\bm{h}(\hat{\alpha}, \hat{\beta})$.

\subsubsection{Direct Estimation of CIR}
Directly estimating the samples of the CIR is a common strategy for channel estimation~\cite{jamali_channel_2016}.
In particular, we use a least-squares scheme solving for $\bm{h}$:
\begin{equation}
    \label{eq:LS fit CIR}
        \hat{\bm{h}} = \argmin_{\bm{h}\in\mathbb{R}^{N I}} \,\lVert \rtrain - \bm{s}(\atrain,\bm{h})\rVert^2.
\end{equation}
This is a linear least-squares optimization problem with $N I$ variables.
Typically, $N I>3$, i.e., a larger number of variables has to be estimated compared to the parametric approach in \eqref{eq:nonlinear estimation}.
In the numerical results in Section~\ref{sec:experiments}, we compare the performance of both methods.
\changed{For the numerical solution of \eqref{eq:nonlinear estimation} and \eqref{eq:LS fit CIR}, we use standard solvers from the Matlab curve fitting toolbox, which are based on gradient descent.}

\subsection{Detection Algorithms}
\label{sec:detection algorithms}
For detection, i.e., estimation of the transmitted bit sequence from the received signal, we consider both sequence estimation assuming the PAM structure in \eqref{eq:PAM model discrete} as well as a model-agnostic heuristic detection scheme, namely increase detection.

\subsubsection{Sequence Estimation}
For sequence estimation, we employ \changed{a standard} Viterbi algorithm which solves the following optimization problem \cite[Chapter 10]{proakis_digital_2001}
\begin{equation}
\label{eq:LS fit symbols}
\hat{\ainfo} = \argmin_{\ainfo\in\{0,1\}^Ki} \, \lVert \bm{r} - \bm{s}([\atrain\,\ainfo], \hat{\bm{h}})\rVert^2,
\end{equation}
where $\ainfo$ is the sequence of information symbols of length $\Ki$ and $\hat{\bm{h}}$ can be either estimated based on our proposed model via \eqref{eq:nonlinear estimation} or directly via \eqref{eq:LS fit CIR}.
We note that this criterion is optimal with respect to the error rate in case of impairment by additive white Gaussian noise \cite{proakis_digital_2001} but is not necessarily optimal for the unknown distortions in our testbed.
The performance in terms of the number of decision errors achievable with sequence estimation is evaluated in Section~\ref{sec:experiments}.

\subsubsection{Increase Detection}
\changed{Since the sequence estimation described before can be computationally demanding and is sensitive to model mismatches, in the following, we consider a low-complexity model-agnostic baseline scheme.}
In particular, this detection method does not rely on the PAM model introduced above.
Instead, it is a heuristic attempt to exploit the observed characteristics of the received signal.
In particular, in a given symbol interval, for a binary ``1'' the received signal exhibits an increase in the current symbol interval (after an appropriate delay) whereas for a binary ``0'' the received signal is non-increasing on average.
This appears to be a convenient signal characteristic to exploit for detection when the exact channel distortions are unknown.
Hence, one heuristic approach for detection is, for each symbol interval and despite the ISI, to check whether the signal is significantly increasing or not.
In particular, we employ for the estimated symbol sequence $\hat{\bm{a}}$ the following detection rule
\begin{equation}
\label{eq:increase detection}
\hat{a}[k] =
\begin{cases}
1, & \mathrm{if}\quad r[i_2[k]] - r[i_1[k]] > \xi \\
0, & \mathrm{otherwise}
\end{cases}
\end{equation}
where $i_1[k] = kI, k\in[0,K)$ is the starting time of the $k$th symbol interval and $i_2[k] = i_1[k] + \sampoff$, where $\sampoff\in[0,I)$ is a sampling offset.
Moreover, $\xi$ is the detection threshold which needs to be carefully selected to balance sensitivity to noise (if $\xi$ is too small) and a bias for detecting binary ``0'' (if $\xi$ is too large).
\changed{We note that the detection rule in \eqref{eq:increase detection} is a common and basic approach for detection and can be seen as a special case of the detection scheme described in \cite{li_lowcomplexity_2016} and as a generalization of the ones in \cite{farsad_tabletop_2013,damrath_low_2016}.}

For choosing the detection parameters $\sampoff$ and $\xi$, there are different options.
In this paper, we obtain these parameters based on peak position $\tpeak$ in \eqref{eq:tpeak} and peak height $\hpeak$ in \eqref{eq:hpeak} of the proposed CIR equation in \eqref{eq:full model} where, for simplicity, we assume that $\alpha=\beta=3$ yields a reasonable characterization of the CIR.
In particular, we choose $\xi=\hpeak/20$ and $\sampoff=\min\{I-1,\ipeak\}$, where $\ipeak=\lceil(\tpeak-\tstart)/\dt\rfloor$ with $\lceil\cdot\rfloor$ denoting rounding to the closest integer value, i.e., we determine the index of the expected peak position without accounting for ISI and not exceeding the symbol interval length.
We note that this detection rule can be seen as a generalization to the one employed in \cite{farsad_tabletop_2013} where $i_1[k]$ is fixed to the middle of the $k$th symbol interval and $i_2[k]$ is fixed to the end of the $k$th symbol interval.


\section{Experimental Results}
\label{sec:experiments}
In this section, we evaluate the analytical model equations in Section~\ref{sec:mathematical model} and fit the parameters of the analytical model to experimental measurement data for the CIR.
Then, we evaluate the performance of the proposed detection schemes.
In the following, the parameter values provided in Table~\ref{tab:system parameters} apply unless indicated otherwise.

\subsection{CIR Model Evaluation}
\label{sec:cir model evaluation}

\begin{figure*}[!t]
    \centering
    \includegraphics[]{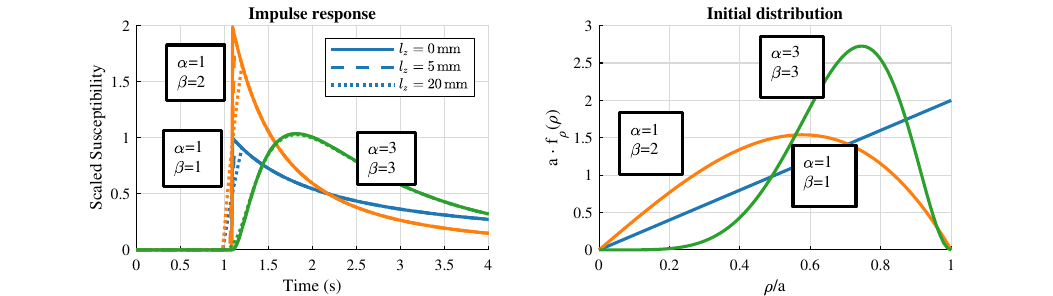}
    \caption{
        Dependence of the CIR on different initial distributions.
        (Left) CIR for distance $d=\SI{10}{\centi\meter}$ scaled with $\myscal$ in \eqref{eq:spawc model} for different Beta initial distribution parameters and receiver weighting function lengths.
        (Right) Corresponding model initial distribution $f_\rho(\rho)$.
        CIRs strongly depend on the particle distribution, e.g., higher CIR peaks and faster decays are observed when relatively more particles are concentrated in the center of the tube rather than at the boundary ($\rho=a$).
    }
    \label{fig:ir init}
\end{figure*}

To better illustrate the properties of the proposed CIR model, we investigate the dependence of the CIR model equation \eqref{eq:spawc model} on the initial particle distribution as well as the impact of the weighting function lengths. 
To this end, in Fig.~\ref{fig:ir init}, we show (left) the numerical evaluation of the CIR and (right) the corresponding initial particle distributions in terms of the radial distribution $f_\rho(\rho)$ in \eqref{eq:fs spec}.
For each initial distribution, CIRs are shown for a receiver length of $l_z=\SI{20}{\milli\meter}$ corresponding to the length of the susceptometer housing (see Fig.~\ref{fig:photo}), $l_z=\SI{5}{\milli\meter}$ corresponding to the sensitive region specified in the manual of the susceptometer, and $l_z=\SI{0}{\milli\meter}$ which can be seen as an approximation and for which the closed-form expression is given in \eqref{eq:full model}.
For the Beta initial distribution, we consider the parameter pairs $(\alpha=1, \beta=1)$ corresponding to a uniform distribution, $(\alpha=1,\beta=2)$ corresponding to a distribution proportional to the flow profile as introduced in Section~\ref{sec:tx spec}, and $(\alpha=3,\beta=3)$ which is chosen arbitrarily.
%
%
For all shown CIRs, we observe an initial delay of about $\SI{1}{\second}$ which is in good agreement with the starting time $t_0=\SI{1.09}{\second}$ in \eqref{eq:tstart}.
More accurately, the signals start at time $(d-l_z/2)/\vmax$ because the receiver weighting function is centered at $z=0$ and extends to $z=-l_z/2$, see Fig.~\ref{fig:abstract system}.

Overall, the observed CIR shapes depend strongly on the initial particle distribution but less on the receiver weighting function length.
Nevertheless, the CIR shapes seem more affected by the choice of different weighting function lengths in case of the uniform distribution and the distribution proportional to the flow profile.
This is in particular the case for the peak value which, in this case ($\alpha=1$), coincides with the signal starting time (see \eqref{eq:tpeak}) and can be attributed to the significant portion of particles around $\rho=0$, see Fig.~\ref{fig:ir init} (right).
This portion of particles travels in a relatively concentrated manner due to the flat flow profile around $\rho=0$, see Fig.~\ref{fig:abstract system}.
Hence, the integral over space in \eqref{eq:first susceptibility model} depends more strongly on the window length.
This is in contrast to the CIR for $(\alpha=3,\beta=3)$ where the CIR depends on the window function length less strongly.
In that case, the initial CIR increase is more smoothly and the approximate peak time occurs at $t=\SI{1.82}{\second}$ via \eqref{eq:tpeak}.

%

In summary, a variety of different CIR shapes can be realized by considering different initial particle distributions.
However, only minor variations in the CIR shape are observed for the considered different weighting function lengths.
This is especially true for the distributions with diminishing mass at $\rho=0$ that are expected for the presented testbed, see also Fig.~\ref{fig:injection} where most of the visible particle cloud resides within the upper half of the tube.
Hence, in the following, the approximation of zero window length in \eqref{eq:full model} is used for fitting of the measurement data due to its mathematical simplicity.
Nevertheless, the more general CIR expression in \eqref{eq:spawc model} might be convenient when investigating the effect of different coil lengths for a custom susceptometer.

\subsection{Conducted Experiments}
\label{sec:conducted experiments}
To test the applicability of our analytical model, we make the following two experiments.

\paragraph{Pulse Train}
For this experiment, we transmit a fixed sequence of 15 binary ``1'' via OOK with symbol durations $T=\SIlist{20;40;60;60}{\second}$ for distances of $d=\SIlist{5;10;20;40}{\centi\meter}$ such that by visual inspection no ISI is present, i.e., we interpret the observed consecutive pulses as realizations of the CIR.
From this data, we can then obtain a measured average CIR and also evaluate variations of the CIR.
%

\paragraph{Data Transmission}
For this experiment, we transmit a fixed sequence of 400 randomly (for time synchronization, the first symbol is fixed to be a ``1'') chosen binary OOK symbols with symbol duration $T=\SI{1}{\second}$ for distances $d=\SIlist{5;10;20;40}{\centi\meter}$, i.e., we take ISI into account.
From this data, we obtain estimates of the CIR using both \eqref{eq:nonlinear estimation} and \eqref{eq:LS fit CIR} and perform detection using both sequence estimation and increase detection.

\subsection{CIR Estimation}
In this subsection, we evaluate the channel estimation scheme as described in Section~\ref{sec:channel estimation} visually and in terms of the root mean square error (RMSE).

\begin{figure*}[!t]
    \centering
    \includegraphics[]{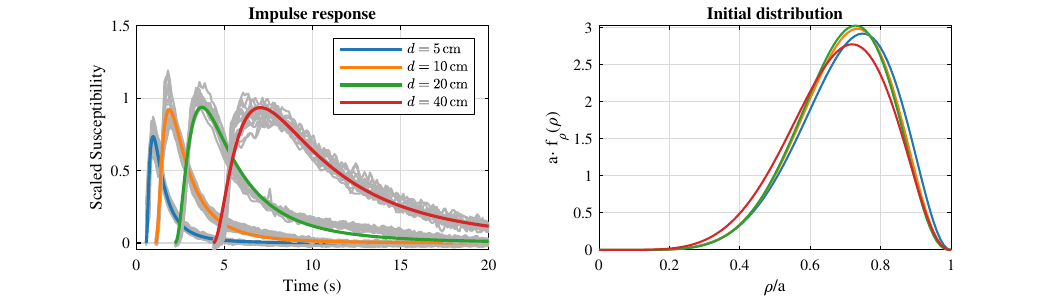}
    \caption{
        %
        %
        %
        %
        Fitting of CIR model.
        (Left) 15 overlayed CIR realizations (gray curves) and estimated CIR model (blue, orange, green, and red curves) for distances of $d=\SIlist{5;10;20;40}{\centi\meter}$.
        The CIRs are scaled by $\myscal$ and shifted to time $\tstart$.
        There is a good fit between model and measurement.
        (Right) Corresponding fitted initial distributions.
        The fitting parameters are $(\alpha=3.41,\beta=3.28,\scal=0.69), (\alpha=3.59,\beta=3.65,\scal=0.81), (\alpha=3.70,\beta=3.83,\scal=0.80), (\alpha=3.13,\beta=3.47,\scal=0.81)$.
        The fitted initial distributions are consistent across all considered distances.
    }
    \label{fig:fitted model}
\end{figure*}
%
%
%
%
In Fig.~\ref{fig:fitted model}, we evaluate the channel estimation based on our proposed CIR model in \eqref{eq:nonlinear estimation} for the pulse train experiments described in Section~\ref{sec:conducted experiments}.
Thereby, we consider CIR lengths of \numlist{10;15;20;20} symbol durations (excluding the initial delay) for distances $d=\SIlist{5;10;20;40}{\centi\meter}$.
On the left hand side, for each transmission distance, we show an overlay of 15 CIR measured pulses (gray curves) as well as the estimated CIR with fitted parameters (blue, orange, green, and red curves) according to \eqref{eq:nonlinear estimation}.
For illustration, the synchronized CIRs are shifted to start consistently at time $\tstart=d/\vmax$ and all CIRs are scaled by $\myscal$.
On the right hand side, we show the corresponding fitted initial particle distributions.
%
%
%
%

From the CIR data (left), we can observe that the measured CIRs do not show much variation across the considered 15 realizations.
Furthermore, the peak height for the scaled CIRs is similar for all considered transmission distances $d$ which means that the peak heights of the unscaled CIRs scale approximately as $1/d$.
Moreover, the CIRs become significantly broader for increasing distance which can be associated with increasing levels of ISI.
From the fitted initial particle distributions (right), we can observe a similarity for all considered distances which is consistent with our model since the initial distribution is assumed to depend on the injection but not on the transmission distance.
The initial radial particle distributions exhibit a peak around $\rho=0.75a$ and diminishing mass at $\rho=0$ and $\rho=a$.
%
%
%
%

In summary, the derived CIR model can fit measurement data remarkably well despite the underlying simplifying assumptions and is consistent in terms of peak decay over distance and stable in the initial particle distribution for different distances.
Moreover, the measured individual CIRs are not significantly affected by noise or other distortions, i.e., the randomness of the CIR is limited for the considered operation of the testbed.
In the following, we investigate how the proposed CIR model generalizes to the data transmission experiments described in Section~\ref{sec:conducted experiments}.
To this end, we perform CIR estimation on the first $\Ktrain$ symbols and then evaluate the root mean square error (RMSE) for all 400 symbols $\bm{a}$.
The RMSE is normalized per sample and can be computed as $\mathrm{RMSE}=\sqrt{|\bm{r} -\bm{s}(\bm{a},\scal\cdot\bm{h})|^2/400/\myscal^2}$.
Thereby, for obtaining $\bm{h}$ both the model-based estimate \eqref{eq:nonlinear estimation} as well as the sample-based estimate \eqref{eq:LS fit CIR} in Section~\ref{sec:channel estimation} are evaluated.

\begin{figure}
    \centering
    \includegraphics{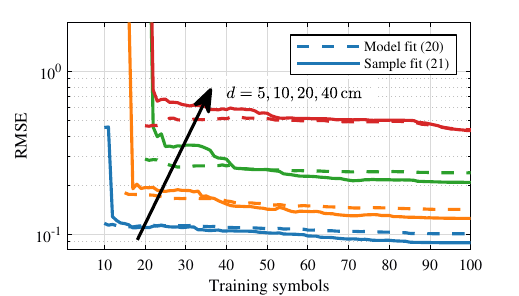}
    \caption{
    Quality of channel estimation.
    The average RMSE per sample for the whole received signal is shown as a function of the number of training symbols used for channel estimation for distances $d=\SIlist{5;10;20;40}{\centi\meter}$.
    The received signals are scaled by $\myscal$.
    Channel estimation by fitting of the proposed model and by fitting of all CIR samples are compared.
    The CIR estimate using the proposed model is reasonably accurate for all considered numbers of training symbols.
    }
    \label{fig:rmse}
\end{figure}
The corresponding results for distances $d=\SIlist{5;10;20;40}{\centi\meter}$ are shown in Fig.~\ref{fig:rmse}, where the RMSE for the information symbols is shown as a function of the number of training symbols used for CIR estimation.
\changed{The RMSE values are generally in the order of \num{e-1} which is reasonable considering the normalized signal values in Fig.~\ref{fig:fitted model} and a slight model mismatch.}
For increasing distance, we generally have larger errors and the RMSE generally decreases for more training symbols, i.e., the estimates are more accurate if the training is longer but generally worse for more ISI.
This mismatch could be caused by several physical effects related to the injection or reception and is worthwhile to study in future work.
The model-based estimate works reasonably well for all considered numbers of training symbols, i.e., the estimate generalizes well even for small numbers of training symbols.
The sample-based estimate strongly depends on the number of training symbols and improves as the number of training symbols increases.
Thereby, the model-based estimate outperforms the sample-based estimate for smaller numbers of training symbols while the sample-based estimate can be slightly better for long training.

\subsection{Detection}
To investigate the performance of the proposed detection schemes, we apply increase detection and sequence estimation as described in Section~\ref{sec:demodulation} for detection of the 400 OOK symbol sequence transmitted in our data transmission experiments described in Section~\ref{sec:conducted experiments}.
Thereby, based on the RMSE results in Fig.~\ref{fig:rmse}, we choose the training as follows.
On the one hand, for CIR estimation using \eqref{eq:nonlinear estimation}, the first $\Ktrain=\numlist{10;15;20;20}$ symbols are used as training symbols, as the RMSE does not significantly decrease for longer training sequences.
On the other hand, for CIR estimation using \eqref{eq:LS fit CIR}, the first $\Ktrain=\numlist{50;75;100;100}$ symbols are used as training symbols, as the RMSE decreases significantly for larger numbers of training symbols.
The remaining $\Ki=400-\Ktrain$ symbols are used for evaluating the proposed detection algorithms but not for channel estimation.

\begin{table}[!t]
    \caption{
        Number of errors for the last 300 data symbols
    }
    \label{tab:error rates}
    \centering
    \begin{tabular}{lllll}
        \toprule
        Scheme                              &	\SI{5}{\centi\meter}	&	\SI{10}{\centi\meter}	&	\SI{20}{\centi\meter}	&	\SI{40}{\centi\meter}	\\
        \midrule
        Increase Detection                   &	2 	&	0		&	15		&	17		\\
        Sequence Estimation (model)          &	0	&	0		&	11		&	35		\\
        Sequence Estimation (sample)         &	0	&	0		&	0		&	31		\\
        \bottomrule
    \end{tabular}
\end{table}

The corresponding decision error results are summarized in Table~\ref{tab:error rates} where for comparison only errors for the last 300 data symbols are reported.
For distances of up to \SI{10}{\centi\meter}, no or only a small number of decision errors are observed for all considered detection schemes.
For a distance of \SI{20}{\centi\meter}, some symbol errors are observed for both increase detection and sequence estimation with model-based CIR estimation whereas sequence estimation with sample-based CIR estimation still shows no errors.
In this scenario, because of the long training sequence, the sample-based CIR estimate is accurate and hence detection benefits from the more complex sequence estimation algorithm.
For a distance of \SI{40}{\centi\meter}, all considered detection schemes cause decision errors whereby increase detection results in fewer errors than sequence estimation.
In this case, the worse performance of sequence estimation can be attributed to the relatively large CIR estimation error for larger distances, see Fig.~\ref{fig:rmse}.
Increase detection does not rely on CIR estimation, and thus exhibits a similar number of decision errors as for a distance of \SI{20}{\centi\meter}.

In summary, with any of the presented detection schemes, reliable communication with only few decision errors is possible for distances of at least up to \SI{40}{\centi\meter}.
Nevertheless, non-coherent detection schemes like the proposed increase detection might cope better with unknown distortions and non-linearities in cases of severe ISI as is the case for larger distances.
Coherent detection schemes like the proposed sequence estimation are expected to perform well with enough training data where less training is required for the proposed model CIR.
However, we note that the presented results correspond to just a single realization of the received signal for the transmission of 400 symbols and more experiments are necessary to thoroughly evaluate different detection schemes.

\begin{figure*}[!t]
    \centering
    \includegraphics{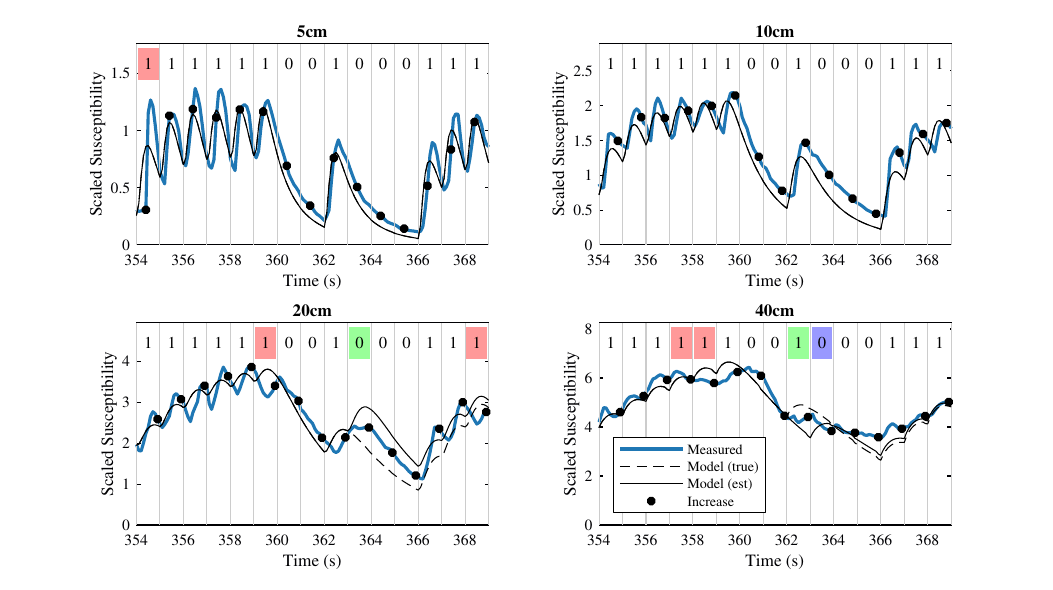}
    \caption{
        %
        %
        %
        %
        %
        %
        Frames of the received signal towards the end of transmission.
        Symbol intervals are separated by gray vertical lines and the transmitted bit sequence is shown as black text.
        Detection errors are indicated by red, blue, and green background for increase detection, sequence estimation, and both, respectively.
        For increase detection, the highlighted black signal points are compared with the signal value at the beginning of each symbol interval.
        For sequence estimation, the model PAM signal using the detected symbol sequence and the transmitted symbol sequence are shown by black solid and dashed lines, respectively.
        The more severe ISI for increasing distances limits the detection performance for larger distances.
    }
    \label{fig:rx signal}
\end{figure*}
%
%

To visualize the model mismatch and to illustrate the detection algorithms and the error events we show excerpts of the received signal in Fig.~\ref{fig:rx signal} for transmission distances of $d=\SIlist{5;10;20;40}{\centi\meter}$.
In particular, for all considered distances, we show the measured received signal scaled by $\myscal$ for a 15 symbol time frame from \SIrange{354}{369}{\second} towards the end of transmission.
Increase detection is visualized by the symbol interval starting times and the signal point used for detection within the interval.
Sequence estimation with the model CIR is visualized by the hypothetical PAM signal using both the estimated symbol sequence as well as the true one.

Generally, as expected from the CIRs in Fig.~\ref{fig:fitted model}, we observe increasingly less pronounced peaks and more ISI for increasing distances.
The model PAM signal appears to follow the measured received signal reasonably well but the mismatch between measurement and model becomes more pronounced for larger distances (note that the y-axes are scaled differently).

The highlighted decision errors can be explained as follows.
For a transmission distance of $d=\SI{5}{\centi\meter}$, for the symbol interval starting at \SI{354}{\second}, a ``1'' to ``0'' error is observed for increase detection which is due to the second signal sample being not significantly larger than the signal sample at the beginning despite there being a significant peak.
This can be explained by a time synchronization error, e.g., shifting the two sample points a little bit to the right the bit ``1'' could be correctly detected.
However, this shift cannot be generally applied because all other symbols would be affected as well, e.g., for the symbol interval starting at \SI{362}{\second} the timing is fine.
For transmission distance \SI{20}{\centi\meter}, for the symbol interval starting at time  \SI{363}{\second}, a ``0'' to ``1'' error \changed{is} observed for both considered detection schemes.
Thereby, for increase detection the error is caused by the second sample being larger than the first sample which might again be caused by small errors introduced by the susceptometer software or noise.
For sequence estimation, the error can be interpreted as a "lift up" of the signal to better follow the measurement at later times, e.g., compare dashed "true" line with solid "detected" line at time \SI{367}{\second}.
The other observed decision errors can be explained in a similar manner.

In summary, the linear PAM model with the estimated CIR is in good agreement with the measurement results.
However, for larger distances, here for \SI{40}{\centi\meter}, strong and potentially non-linear ISI is present.
Moreover, time synchronization and improved receiver concepts constitute interesting topics for future work.

\section{Conclusions}
\label{sec:conclusion}
\subsection{Summary}
In this paper, we have presented a new testbed for the investigation of flow-driven MC systems \changed{where for signal transmission SPIONs (magnetic particles) are injected into a background flow which can be detected downstream by measurement of the magnetic susceptibility without direct access to the tube channel.
The SPIONs are engineered to be chemically stable, to avoid agglomeration, and to not engage in reactions with molecules in the environment.}
After a review of the relevant physical effects, we proposed a simple mathematical model based on laminar flow-driven particle transport, a parametric initial SPION distribution with two parameters, and a transparent receiver.
Channel estimation for several measurements with and without ISI confirmed the applicability of the proposed CIR model for different transmission distances and training sequence lengths.
\changed{Moreover, model-based and model-agnostic symbol detection schemes were shown to enable reliable communication for example measurements.
The presented testbed can contribute to a better understanding of flow-driven MC systems, including the subunits of the cardiovascular system and chemical reactors.}
Potential applications of SPION based MC include reporting sensing results and carrying control information in industrial, microfluidic or biomedical settings, especially at locations where other forms of communication can not be employed.

\subsection{Outlook}
\label{sec:outlook}
We highlight the following directions for future theoretical and experimental work \changed{extending the results presented in this paper.
There are several interesting preprocessing and detection schemes that could be evaluated with the proposed testbed including matched filtering \cite{jamali_design_2017}, optimal coherent and non-coherent \cite{jamali_noncoherent_2018} as well as adaptive, learned \cite{farsad_datadriven_2020}, and feature-based heuristic detection schemes \cite{wei_highdimensional_2020}.}
To this end, it will also be useful to develop further mathematical models for the received signal, including a statistical characterization of noise and other distortions, e.g., by diffusion, turbulent flow, the injection, the properties of the employed fluid, an overall non-linearity, and time-variant flow \cite{jamali_channel_2019}.
Further comprehensive measurements will help in validating these models and algorithms.
These models will also help in developing novel channel estimation \cite{jamali_channel_2016} and synchronization \cite{jamali_symbol_2017} schemes which again can be model-based to different degrees.
In addition to detection, also different modulation schemes and transmission from a single transmitter to multiple receivers as well as from multiple transmitters to a single receiver could be investigated as suitable extensions of the presented point-to-point link.
A better theoretical understanding will also help guiding the hardware development.
This includes the optimization of the receiver device \cite{bartunik_novel_2019}, employing different pumps for better control of the injection, changing the injection mechanism, e.g., replacing the Y-connector by a needle, and testing different types of particles.
Moreover, the testbed could potentially be expanded by implementing a network of ducts, changing the carrier liquid, employing magnets for particle movement control, and scaling of its size.
Furthermore, particles could be additionally tagged with other chemicals.
These extensions could also facilitate the use of higher-order modulation, e.g., by using different particle types, combining optical and magnetic measurements of the particles, or using different forms of injection.

\appendix
\section*{Derivation of Flow-Driven Model Impulse Response}
In this appendix, we derive the model CIR in \eqref{eq:spawc model}.
For the following derivation, we assume cylindrical coordinates $\bm{x}=(\rho,\phi,z)$.

In general, from \eqref{eq:first susceptibility model} and \eqref{eq:c flow}, the received signal due to a single release at time $t=0$ can be written as
\begin{equation}
\label{eq:general impulse response}
\chi(t) = \chiref \cdot \iiint_{\mathbb{R}^3} w(\bm{x})\cdot \cin(\bm{x}-\vel(\rho)\cdot t\cdot \bm{e}_z)\de V.
\end{equation}

Now, using \eqref{eq:rectangular window} and \eqref{eq:cin} and \eqref{eq:frho def}, we arrive at
\begin{equation}
\label{eq:h deriv}
h(t) = \chiref\frac{\Vi}{\pi a^2 l_z}\int_0^a\int_{-\infty}^\infty \rect(z/l_z) \cdot f_\rho(\rho) \cdot\delta(z-\vel(\rho)t+d) \de z \de \rho.
\end{equation}
For convenience, we substitute $\rho$ with $s = (\rho/a)^2$ and use $f_s(s)$ in \eqref{eq:fs spec}.
Then, we obtain
\begin{equation}
h(t) = \chiref\frac{\Vi}{\pi a^2 l_z} \int_0^1 \int_{-\infty}^\infty \rect(z/l_z)\cdot f_s(s) \cdot \delta(\varphi(s)) \de z\de s,
\end{equation}
where $\varphi(s)=z-\tilde{\vel}(s)\cdot t + d$ and $\tilde{\vel}(s) = \vmax\cdot(1-s)$ which is simply obtained from \eqref{eq:flow profile} by substituting $\rho$ with $s$.
Now, we use the properties of the Dirac delta function to simplify the term $\delta(\varphi(s))$.
To this end, we note that $\varphi(s_0)=0$ for $s_0=1-(z+d)/(\vmax t)$ provided $z+d<\tilde{\vel}(s)\cdot t$.
Thus, we can rewrite the delta function as \cite{hoskins_delta_2009}
\begin{equation}
\delta(\varphi(s)) = \frac{1}{|\varphi'(s_0)|} \delta(s-s_0),
\end{equation}
where $\varphi'(s)=\vmax t$.
Then, using the sifting property of the Dirac delta function \cite{hoskins_delta_2009}, we arrive at
\begin{equation}
    \label{eq:derived impulse response}
    h(t) = \chiref\frac{\Vi}{\pi a^2 l_z} \cdot\frac{1}{\vmax t} \int_{-\infty}^\infty \rect(z/l_z) \cdot f_s\left(1 - \frac{z+d}{\vmax t}\right) \de z.
\end{equation}
Finally, by straightforward integration and using the definition of $F_s(s)$, we arrive at \eqref{eq:spawc model}.
This concludes the proof.



\bibliographystyle{IEEEtran}
\bibliography{main.bbl}

\begin{thebibliography}{10}
\providecommand{\url}[1]{#1}
\csname url@samestyle\endcsname
\providecommand{\newblock}{\relax}
\providecommand{\bibinfo}[2]{#2}
\providecommand{\BIBentrySTDinterwordspacing}{\spaceskip=0pt\relax}
\providecommand{\BIBentryALTinterwordstretchfactor}{4}
\providecommand{\BIBentryALTinterwordspacing}{\spaceskip=\fontdimen2\font plus
\BIBentryALTinterwordstretchfactor\fontdimen3\font minus
  \fontdimen4\font\relax}
\providecommand{\BIBforeignlanguage}[2]{{%
\expandafter\ifx\csname l@#1\endcsname\relax
\typeout{** WARNING: IEEEtran.bst: No hyphenation pattern has been}%
\typeout{** loaded for the language `#1'. Using the pattern for}%
\typeout{** the default language instead.}%
\else
\language=\csname l@#1\endcsname
\fi
#2}}
\providecommand{\BIBdecl}{\relax}
\BIBdecl

\bibitem{unterweger_experimental_2018}
H.~Unterweger, J.~Kirchner, W.~Wicke, A.~Ahmadzadeh, D.~Ahmed, V.~Jamali,
  C.~Alexiou, G.~Fischer, and R.~Schober, ``Experimental molecular
  communication testbed based on magnetic nanoparticles in duct flow,'' in
  \emph{Proc. IEEE SPAWC}, Jun. 2018, pp. 1--5.

\bibitem{nakano_molecular_2013}
T.~Nakano, A.~W. Eckford, and T.~Haraguchi,
  \emph{\BIBforeignlanguage{en}{Molecular communication}}.\hskip 1em plus 0.5em
  minus 0.4em\relax {Cambridge University Press}, Sep. 2013.

\bibitem{grebenstein_biological_2019}
L.~Grebenstein, J.~Kirchner, R.~S. Peixoto, W.~Zimmermann, F.~Irnstorfer,
  W.~Wicke, A.~Ahmadzadeh, V.~Jamali, G.~Fischer, R.~Weigel, A.~Burkovski, and
  R.~Schober, ``Biological optical-to-chemical signal conversion interface: A
  small-scale modulator for molecular communications,'' \emph{{IEEE} Trans.
  Nanobiosci.}, vol.~18, no.~1, pp. 31--42, Jan. 2019.

\bibitem{akyildiz_internet_2015}
I.~F. Akyildiz, M.~Pierobon, S.~Balasubramaniam, and Y.~Koucheryavy, ``The
  internet of bio-nano things,'' \emph{{IEEE} Commun. Mag.}, vol.~53, no.~3,
  pp. 32--40, Mar. 2015.

\bibitem{haselmayr_integration_2019}
W.~Haselmayr, A.~Springer, G.~Fischer, C.~Alexiou, H.~Boche, P.~A. Höher,
  F.~Dressler, and R.~Schober, ``\BIBforeignlanguage{en}{Integration of
  molecular communications into future generation wireless networks},'' in
  \emph{\BIBforeignlanguage{en}{Proc. 6G Wireless Summit}}, {Levi, Finland},
  2019, p.~2.

\bibitem{farsad_comprehensive_2016}
N.~Farsad, H.~B. Yilmaz, A.~Eckford, C.~Chae, and W.~Guo, ``A comprehensive
  survey of recent advancements in molecular communication,'' \emph{{IEEE}
  Commun. Surveys Tuts.}, vol.~18, no.~3, pp. 1887--1919, 3rd quart. 2016.

\bibitem{jamali_channel_2019}
V.~Jamali, A.~Ahmadzadeh, W.~Wicke, A.~Noel, and R.~Schober, ``Channel modeling
  for diffusive molecular communication \textemdash{} a tutorial review,''
  \emph{Proc. {IEEE}}, vol. 107, no.~7, pp. 1256--1301, Jul. 2019.

\bibitem{kuscu_transmitter_2019}
M.~Kuscu, E.~Dinc, B.~A. Bilgin, H.~Ramezani, and O.~B. Akan, ``Transmitter and
  receiver architectures for molecular communications: A survey on physical
  design with modulation, coding, and detection techniques,'' \emph{Proc.
  {IEEE}}, vol. 107, no.~7, pp. 1302--1341, Jul. 2019.

\bibitem{farsad_tabletop_2013}
N.~Farsad, W.~Guo, and A.~W. Eckford, ``\BIBforeignlanguage{en}{Tabletop
  molecular communication: Text messages through chemical signals},''
  \emph{\BIBforeignlanguage{en}{PLOS One}}, vol.~8, no.~12, p. e82935, Dec.
  2013.

\bibitem{giannoukos_molecular_2017}
S.~Giannoukos, A.~Marshall, S.~Taylor, and J.~Smith,
  ``\BIBforeignlanguage{en}{Molecular communication over gas stream channels
  using portable mass spectrometry},'' \emph{\BIBforeignlanguage{en}{J. Am.
  Soc. Mass Spectrom.}}, vol.~28, no.~11, pp. 2371--2383, Nov. 2017.

\bibitem{shakya_correlated_2018}
P.~Shakya, E.~Kennedy, C.~Rose, and J.~K. Rosenstein, ``Correlated transmission
  and detection of concentration-modulated chemical vapor plumes,''
  \emph{{IEEE} Sensors J.}, vol.~18, no.~16, pp. 6504--6509, Aug. 2018.

\changed{\bibitem{damrath_investigation_2021}
M.~Damrath, S.~Bhattacharjee, and P.~A. Hoeher, ``Investigation of multiple
  fluorescent dyes in macroscopic air-based molecular communication,''
  \emph{{IEEE} Trans. Mol. Biol. Multi-Scale Commun.}, 2021.}

\bibitem{farsad_novel_2017}
N.~Farsad, D.~Pan, and A.~Goldsmith, ``A novel experimental platform for
  in-vessel multi-chemical molecular communications,'' in \emph{Proc. IEEE
  GLOBECOM}, Dec. 2017, pp. 1--6.

\bibitem{khaloopour_experimental_2019}
L.~Khaloopour, S.~V. Rouzegar, A.~Azizi, A.~Hosseinian, M.~{Farahnak-Ghazani},
  N.~Bagheri, M.~Mirmohseni, H.~Arjmandi, R.~Mosayebi, and M.~{Nasiri-Kenari},
  ``An experimental platform for macro-scale fluidic medium molecular
  communication,'' \emph{{IEEE} Trans. Mol. Biol. Multi-Scale Commun.}, vol.~5,
  no.~3, pp. 163--175, Dec. 2019.

\bibitem{tuccitto_reactive_2018}
N.~Tuccitto, G.~{Li-Destri}, G.~M.~L. Messina, and G.~Marletta,
  ``\BIBforeignlanguage{en}{Reactive messengers for digital molecular
  communication with variable transmitter--receiver distance},''
  \emph{\BIBforeignlanguage{en}{Phys. Chem. Chem. Phys.}}, vol.~20, no.~48, pp.
  30\,312--30\,320, Dec. 2018.

\bibitem{atthanayake_experimental_2018}
I.~Atthanayake, S.~Esfahani, P.~Denissenko, I.~Guymer, P.~J. Thomas, and
  W.~Guo, ``Experimental molecular communications in obstacle rich fluids,'' in
  \emph{Proc. ACM Nanocom}, {Reykjavik, Iceland}, Sep. 2018, pp. 1--2.

\bibitem{koo_deep_2020}
B.-H. Koo, H.~J. Kim, J.-Y. Kwon, and C.-B. Chae, ``Deep learning-based human
  implantable nano molecular communications,'' in \emph{Proc. IEEE ICC}, Jun.
  2020, pp. 1--7.

\bibitem{kuscu_graphenebased_2020}
\BIBentryALTinterwordspacing
M.~Kuscu, H.~Ramezani, E.~Dinc, S.~Akhavan, and O.~B. Akan, ``Graphene-based
  nanoscale molecular communication receiver: Fabrication and microfluidic
  analysis,'' Jul. 2020. [Online]. Available:
  \url{https://arxiv.org/abs/2006.15470}
\BIBentrySTDinterwordspacing

\bibitem{jamali_diffusive_2018}
V.~Jamali, N.~Farsad, R.~Schober, and A.~Goldsmith, ``Diffusive molecular
  communications with reactive molecules: Channel modeling and signal design,''
  \emph{{IEEE} Trans. Mol. Biol. Multi-Scale Commun.}, vol.~4, no.~3, pp.
  171--188, Sep. 2018.

\bibitem{levenspiel_chemical_1999}
O.~Levenspiel, \emph{\BIBforeignlanguage{en}{Chemical reaction
  engineering}}.\hskip 1em plus 0.5em minus 0.4em\relax {Wiley}, 1999.

\bibitem{soldner_survey_2020}
C.~A. S{\"o}ldner, E.~Socher, V.~Jamali, W.~Wicke, A.~Ahmadzadeh, H.-G.
  Breitinger, A.~Burkovski, K.~Castiglione, R.~Schober, and H.~Sticht, ``A
  survey of biological building blocks for synthetic molecular communication
  systems,'' \emph{{IEEE} Commun. Surveys Tuts.}, vol.~22, no.~4, pp. 2765 --
  2800, 4th quart. 2020.

\bibitem{pankhurst_applications_2003}
Q.~A. Pankhurst, J.~Connolly, S.~K. Jones, and J.~Dobson,
  ``\BIBforeignlanguage{en}{Applications of magnetic nanoparticles in
  biomedicine},'' \emph{\BIBforeignlanguage{en}{J. Phys. D: Appl. Phys.}},
  vol.~36, no.~13, pp. R167--R181, Jun. 2003.

\bibitem{lu_magnetic_2007}
A.-H. Lu, E.~L. Salabas, and F.~Sch{\"u}th, ``\BIBforeignlanguage{en}{Magnetic
  nanoparticles: Synthesis, protection, functionalization, and application},''
  \emph{\BIBforeignlanguage{en}{Angew. Chem.-int. Edit.}}, vol.~46, no.~8, pp.
  1222--1244, Feb. 2007.

\bibitem{durr_magnetic_2016}
S.~D{\"u}rr, C.~Bohr, M.~P{\"o}ttler, S.~Lyer, R.~P. Friedrich, R.~Tietze,
  M.~D{\"o}llinger, C.~Alexiou, and C.~Janko,
  ``\BIBforeignlanguage{eng}{Magnetic tissue engineering for voice
  rehabilitation - first steps in a promising field},''
  \emph{\BIBforeignlanguage{eng}{Anticancer Res.}}, vol.~36, no.~6, pp.
  3085--3091, Jun. 2016.

\bibitem{giouroudi_microfluidic_2013}
I.~Giouroudi and F.~Keplinger, ``\BIBforeignlanguage{eng}{Microfluidic
  biosensing systems using magnetic nanoparticles},''
  \emph{\BIBforeignlanguage{eng}{Int. J. Mol. Sci.}}, vol.~14, no.~9, pp.
  18\,535--18\,556, Sep. 2013.

\bibitem{gleich_tomographic_2005}
B.~Gleich and J.~Weizenecker, ``\BIBforeignlanguage{eng}{Tomographic imaging
  using the nonlinear response of magnetic particles},''
  \emph{\BIBforeignlanguage{eng}{Nature}}, vol. 435, no. 7046, pp. 1214--1217,
  Jun. 2005.

\bibitem{dobson_remote_2008}
J.~Dobson, ``\BIBforeignlanguage{eng}{Remote control of cellular behaviour with
  magnetic nanoparticles},'' \emph{\BIBforeignlanguage{eng}{Nat.
  Nanotechnol.}}, vol.~3, no.~3, pp. 139--143, Mar. 2008.

\bibitem{raj_coconut_2015}
K.~G. Raj and P.~A. Joy, ``\BIBforeignlanguage{en}{Coconut shell based
  activated carbon--iron oxide magnetic nanocomposite for fast and efficient
  removal of oil spills},'' \emph{\BIBforeignlanguage{en}{J. Environ. Chem.
  Eng.}}, vol.~3, no.~3, pp. 2068--2075, Sep. 2015.

\bibitem{tietze_efficient_2013}
R.~Tietze, S.~Lyer, S.~D{\"u}rr, T.~Struffert, T.~Engelhorn, M.~Schwarz,
  E.~Eckert, T.~G{\"o}en, S.~Vasylyev, W.~Peukert, F.~Wiekhorst, L.~Trahms,
  A.~D{\"o}rfler, and C.~Alexiou, ``\BIBforeignlanguage{eng}{Efficient
  drug-delivery using magnetic nanoparticles -- biodistribution and therapeutic
  effects in tumour bearing rabbits},''
  \emph{\BIBforeignlanguage{eng}{Nanomed.}}, vol.~9, no.~7, pp. 961--971, Oct.
  2013.

\bibitem{wicke_magnetic_2019}
W.~Wicke, A.~Ahmadzadeh, V.~Jamali, H.~Unterweger, C.~Alexiou, and R.~Schober,
  ``Magnetic nanoparticle-based molecular communication in microfluidic
  environments,'' \emph{{IEEE} Trans. Nanobiosci.}, vol.~18, no.~2, pp.
  156--169, Apr. 2019.

\bibitem{schafer_nd_2018}
M.~Sch{\"a}fer, W.~Wicke, R.~Rabenstein, and R.~Schober, ``An {{nD}} model for
  a cylindrical diffusion-advection problem with an orthogonal force
  component,'' in \emph{Proc. IEEE DSP Conf.}, Nov. 2018, pp. 1--5.

\bibitem{schafer_analytical_2019}
------, ``Analytical models for particle diffusion and flow in a horizontal
  cylinder with a vertical force,'' in \emph{Proc. IEEE ICC}, May 2019, pp.
  1--7.

\bibitem{schafer_transfer_2020}
\BIBentryALTinterwordspacing
M.~Sch{\"a}fer, W.~Wicke, L.~Brand, R.~Rabenstein, and R.~Schober, ``Transfer
  function models for cylindrical {{MC}} channels with diffusion and laminar
  flow,'' \emph{submitted for publication}, Jul. 2020. [Online]. Available:
  \url{https://arxiv.org/abs/2007.01799}
\BIBentrySTDinterwordspacing

\bibitem{kisseleff_magnetic_2017}
S.~Kisseleff, R.~Schober, and W.~H. Gerstacker, ``Magnetic nanoparticle based
  interface for molecular communication systems,'' \emph{{IEEE} Commun. Lett.},
  vol.~21, no.~2, pp. 258--261, Feb. 2017.

\bibitem{bartunik_novel_2019}
M.~Bartunik, M.~L{\"u}bke, H.~Unterweger, C.~Alexiou, S.~Meyer, D.~Ahmed,
  G.~Fischer, W.~Wicke, V.~Jamali, R.~Schober, and J.~Kirchner, ``Novel
  receiver for superparamagnetic iron oxide nanoparticles in a molecular
  communication setting,'' in \emph{Proc. ACM Nanocom}, {New York, NY, USA},
  2019, pp. 27:1--27:6.

\bibitem{ahmed_characterization_2019}
D.~Ahmed, H.~Unterweger, G.~Fischer, R.~Schober, and J.~Kirchner,
  ``Characterization of an inductance-based detector in molecular communication
  testbed based on superparamagnetic iron oxide nanoparticles,'' in \emph{Proc.
  IEEE Sensors}, Oct. 2019, pp. 1--4.

\bibitem{bartunik_comparative_2020}
M.~Bartunik, H.~Unterweger, C.~Alexiou, R.~Schober, M.~L{\"u}bke, G.~Fischer,
  and J.~Kirchner, ``\BIBforeignlanguage{en}{Comparative evaluation of a new
  sensor for superparamagnetic iron oxide nanoparticles in a molecular
  communication setting},'' in \emph{\BIBforeignlanguage{en}{Proc. EAI BICT}},
  2020, pp. 303--316.

\bibitem{bartunik_amplitude_2020}
M.~Bartunik, T.~Thalhofer, C.~Wald, M.~Richter, G.~Fischer, and J.~Kirchner,
  ``\BIBforeignlanguage{en}{Amplitude modulation in a molecular communication
  testbed with superparamagnetic iron oxide nanoparticles and a micropump},''
  in \emph{\BIBforeignlanguage{en}{Proc. EAI BodyNets}}, 2020, pp. 92--105.

\bibitem{schlechtweg_magnetic_2019}
N.~Schlechtweg, S.~Meyer, H.~Unterweger, M.~Bartunik, D.~Ahmed, W.~Wicke,
  V.~Jamali, C.~Alexiou, G.~Fischer, R.~Weigel, R.~Schober, and J.~Kirchner,
  ``\BIBforeignlanguage{en}{Magnetic steering of superparamagnetic
  nanoparticles in duct flow for molecular communication: A feasibility
  study},'' in \emph{\BIBforeignlanguage{en}{Proc. EAI BodyNets}}, 2019, pp.
  161--174.

\bibitem{noel_channel_2016}
\BIBentryALTinterwordspacing
A.~Noel, D.~Makrakis, and A.~Hafid, ``Channel impulse responses in diffusive
  molecular communication with spherical transmitters,'' in \emph{Proc. CSIT
  BSC}, Jun. 2016. [Online]. Available: \url{https://arxiv.org/abs/1604.04684}
\BIBentrySTDinterwordspacing

\bibitem{massart_preparation_1981}
R.~Massart, ``Preparation of aqueous magnetic liquids in alkaline and acidic
  media,'' \emph{{IEEE} Trans. Magn.}, vol.~17, no.~2, pp. 1247--1248, Mar.
  1981.

\bibitem{khalafalla_preparation_1980}
S.~Khalafalla and G.~Reimers, ``Preparation of dilution-stable aqueous magnetic
  fluids,'' \emph{{IEEE} Trans. Magn.}, vol.~16, no.~2, pp. 178--183, Mar.
  1980.

\bibitem{zaloga_development_2014}
J.~Zaloga, C.~Janko, J.~Nowak, J.~Matuszak, S.~Knaup, D.~Eberbeck, R.~Tietze,
  H.~Unterweger, R.~P. Friedrich, S.~Duerr, R.~{Heimke-Brinck}, E.~Baum,
  I.~Cicha, F.~D{\"o}rje, S.~Odenbach, S.~Lyer, G.~Lee, and C.~Alexiou,
  ``Development of a lauric acid/albumin hybrid iron oxide nanoparticle system
  with improved biocompatibility,'' \emph{Int. J. Nanomed.}, vol.~9, pp.
  4847--4866, Oct. 2014.

\bibitem{coey_magnetism_2010}
J.~M.~D. Coey, \emph{\BIBforeignlanguage{en}{Magnetism and magnetic
  materials}}.\hskip 1em plus 0.5em minus 0.4em\relax {Cambridge University
  Press}, Mar. 2010.

\bibitem{deen_introduction_2016}
W.~M. Deen, \emph{\BIBforeignlanguage{en}{Introduction to chemical engineering
  fluid mechanics}}.\hskip 1em plus 0.5em minus 0.4em\relax {Cambridge
  University Press}, Aug. 2016.

\bibitem{drees_efficient_2020}
J.~P. Drees, L.~Stratmann, F.~Bronner, M.~Bartunik, J.~Kirchner, H.~Unterweger,
  and F.~Dressler, ``Efficient simulation of macroscopic molecular
  communication: The pogona simulator,'' in \emph{Proc. ACM Nanocom}, New York,
  NY, USA, Sep. 2020.

\bibitem{papoulis_probability_2002}
A.~Papoulis and S.~U. Pillai, \emph{\BIBforeignlanguage{en}{Probability, random
  variables, and stochastic processes}}.\hskip 1em plus 0.5em minus 0.4em\relax
  {McGraw-Hill}, 2002.

\bibitem{wicke_modeling_2018}
W.~Wicke, T.~Schwering, A.~Ahmadzadeh, V.~Jamali, A.~Noel, and R.~Schober,
  ``Modeling duct flow for molecular communication,'' in \emph{Proc. IEEE
  GLOBECOM}, Dec. 2018, pp. 206--212.

\bibitem{jamali_channel_2016}
V.~Jamali, A.~Ahmadzadeh, C.~Jardin, H.~Sticht, and R.~Schober, ``Channel
  estimation for diffusive molecular communications,'' \emph{{IEEE} Trans.
  Commun.}, vol.~64, no.~10, pp. 4238--4252, Oct. 2016.

\bibitem{proakis_digital_2001}
J.~G. Proakis, \emph{\BIBforeignlanguage{en}{Digital communications}}.\hskip
  1em plus 0.5em minus 0.4em\relax {McGraw-Hill}, 2001.

\bibitem{li_lowcomplexity_2016}
B.~Li, M.~Sun, S.~Wang, W.~Guo, and C.~Zhao, ``Low-complexity noncoherent
  signal detection for nanoscale molecular communications,'' \emph{{IEEE}
  Trans. Nanobiosci.}, vol.~15, no.~1, pp. 3--10, Jan. 2016.

\changed{\bibitem{damrath_low_2016}
M.~Damrath and P.~A. Hoeher, ``Low-complexity adaptive threshold detection for
  molecular communication,'' \emph{{IEEE} Trans. Nanobiosci.}, vol.~15, no.~3,
  pp. 200--208, 2016.}

\bibitem{jamali_design_2017}
V.~Jamali, A.~Ahmadzadeh, and R.~Schober, ``On the design of matched filters
  for molecule counting receivers,'' \emph{{IEEE} Commun. Lett.}, vol.~21,
  no.~8, pp. 1711--1714, Aug. 2017.

\bibitem{jamali_noncoherent_2018}
V.~Jamali, N.~Farsad, R.~Schober, and A.~Goldsmith, ``Non-coherent detection
  for diffusive molecular communication systems,'' \emph{{IEEE} Trans.
  Commun.}, vol.~66, no.~6, pp. 2515--2531, Jun. 2018.

\bibitem{farsad_datadriven_2020}
\BIBentryALTinterwordspacing
N.~Farsad, N.~Shlezinger, A.~J. Goldsmith, and Y.~C. Eldar, ``Data-driven
  symbol detection via model-based machine learning,'' Feb. 2020. [Online].
  Available: \url{https://arxiv.org/abs/2002.07806}
\BIBentrySTDinterwordspacing

\bibitem{wei_highdimensional_2020}
Z.~Wei, W.~Guo, B.~Li, J.~Charmet, and C.~Zhao, ``High-dimensional metric
  combining for non-coherent molecular signal detection,'' \emph{{IEEE} Trans.
  Commun.}, vol.~68, no.~3, pp. 1479--1493, Mar. 2020.

\bibitem{jamali_symbol_2017}
V.~Jamali, A.~Ahmadzadeh, and R.~Schober, ``Symbol synchronization for
  diffusion-based molecular communications,'' \emph{{IEEE} Trans. Nanobiosci.},
  vol.~16, no.~8, pp. 873--887, Dec. 2017.

\bibitem{hoskins_delta_2009}
R.~F. Hoskins, \emph{\BIBforeignlanguage{en}{Delta functions: Introduction to
  Generalised Functions}}.\hskip 1em plus 0.5em minus 0.4em\relax {Elsevier},
  Mar. 2009.

\end{thebibliography}

\begin{IEEEbiography}
	[{\includegraphics[width=1in,height=1.25in,clip,keepaspectratio]{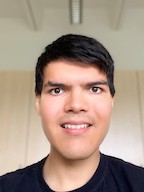}}]{Wayan Wicke} (S'17) was born in Nuremberg, Germany, in 1991.
    He received the B.Sc.\ and  M.Sc.\ degrees in electrical engineering from the Friedrich-Alexander University Erlangen-Nürnberg (FAU), Erlangen, Germany, in 2014 and 2017, respectively, where he is currently pursuing the Ph.D.\ degree.
    His research interests include statistical signal processing and digital communications with a focus on molecular communication.
\end{IEEEbiography}

\begin{IEEEbiography}
    [{\includegraphics[width=1in,height=1.25in,clip,keepaspectratio]{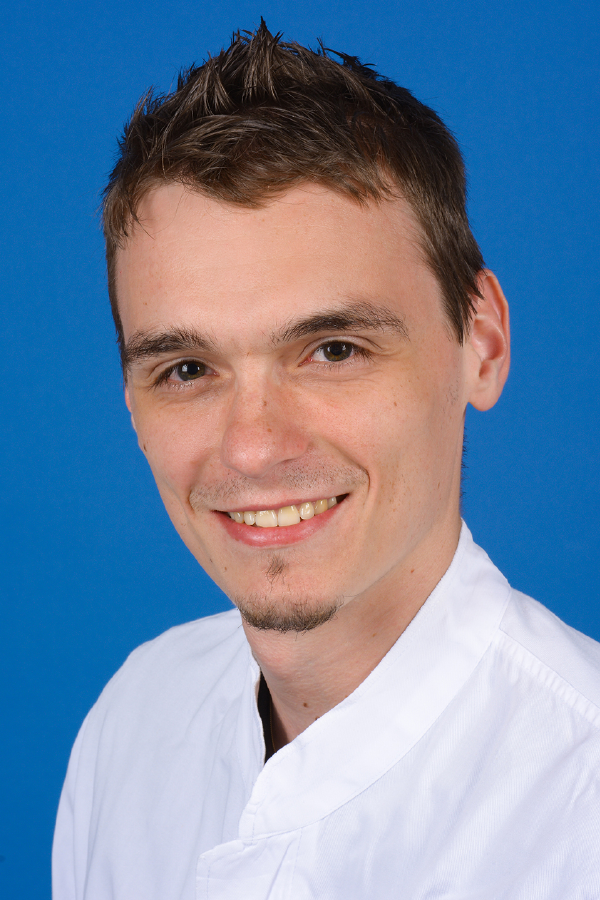}}]
    {Harald Unterweger}
    is a PostDoc and deputy head of the synthesis and analytics department of the Section of Experimental Oncology and Nanomedicine (SEON).
    His work focuses in the development and characterization of magnetic nanoparticles for biomedical and technical applications.
    Harald earned a M.Sc.\ degree in nanotechnology and a Ph.D.\ degree in material sciences from the Friedrich-Alexander University Erlangen-Nürnberg (FAU), Erlangen, Germany.
    For his Ph.D.\ thesis, he received the dissertation award from the German Ferrofluid Society and the dissertation award from the FAU's Technical Faculty (Freundeskreis der Alumni Technische Fakult\"at Erlangen).
\end{IEEEbiography}

\begin{IEEEbiography}[{\includegraphics[width=1in,height=1.25in,clip,keepaspectratio]{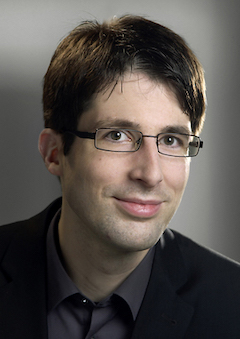}}]{Jens Kirchner} (Senior Member, IEEE) studied physics at Friedrich-Alexander-Universität Erlangen-Nürnberg (FAU), Germany, and University of St. Andrews, Scotland, and received the diploma degree from FAU in 2004. He received his doctorates in 2008 and 2016 from FAU in the fields of biosignal analysis and philosophy of science, respectively. Between 2008 and 2015 he worked at Biotronik SE \& Co. KG in Erlangen and Berlin in the research and development of implantable cardiac sensors. In 2015, he joined the Institute for Electronics Engineering at FAU, where he heads the Medical Electronics \& Multiphysics Systems group. His research interests lie in wearable and implantable sensors, inductive power transfer, and molecular communication.
\end{IEEEbiography}

\begin{IEEEbiography}[{\includegraphics[width=1in,height=1.25in,clip,keepaspectratio]{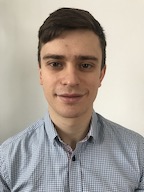}}]{Lukas Brand} received the B.Sc. degree in electrical engineering, and the M.Sc. degree in advanced signal processing and communications engineering, an Elite Master's programme within the Elite Network of Bavaria, from the Friedrich-Alexander-Universität (FAU), Erlangen, Germany, in 2017 and 2019, respectively, where he is currently working towards the Ph.D. degree in electrical engineering at the Institute for Digital Communications. His current research interests include wireless and molecular communications.
\end{IEEEbiography}

\begin{IEEEbiography}
	[{\includegraphics[width=1in,height=1.25in,clip,keepaspectratio]{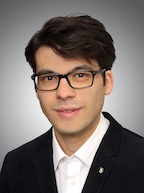}}]{Arman Ahmadzadeh} (Member, IEEE) received
the B.Sc. degree in electrical engineering from the
Ferdowsi University of Mashhad, Mashhad, Iran,
in 2010, and the M.Sc. degree in communications
and multimedia engineering from the Friedrich-Alexander-Universität (FAU) Erlangen-Nürnberg,
Erlangen, Germany, in 2013, where he is currently pursuing the Ph.D. degree in electrical engineering with the Institute for Digital Communications. His current research interest includes physical layer
molecular communications. He has served as a member of the Technical Program Committee of the Communication Theory Symposium for the IEEE International Conference on Communications (ICC) from
2017 to 2020. He received several awards, including the Best Paper Award
from the IEEE ICC in 2016, IEEE ICC in 2020, and the Student Travel Grants
for attending the Global Communications Conference (GLOBECOM) in 2017.
He was recognized as an Exemplary Reviewer of IEEE COMMUNICATIONS
LETTERS in 2016
\end{IEEEbiography}

\begin{IEEEbiography}[{\includegraphics[width=1in,height=1.25in,clip,keepaspectratio]{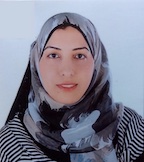}}]{Doaa Ahmed} was born in Ismailia, Egypt in 1988. She received the B.Sc. degree (with Honors) in Communications and Electronics Engineering from Suez Canal University, Egypt, in 2010 and M.Sc. degree in Communications and Multimedia Engineering from Friedrich-Alexander University Erlangen-Nuremberg, Erlangen, Germany, in 2016.
From 2010 to 2014, she worked as a Research and Teaching Assistant with the Chair of Communications and Electronics, Suez Canal University, Egypt. 
She was a reviewer in the IEEE-EMBS International Conference 
on Biomedical and Health Informatics (BHI) for the years 2019 and 2021.
She is currently pursuing the Ph.D. degree at Friedrich-Alexander University Erlangen-Nuremberg, Erlangen, Germany.
\end{IEEEbiography}

\begin{IEEEbiography}[{\includegraphics[width=1in,height=1.25in,clip,keepaspectratio]{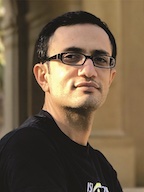}}]{Vahid Jamali} (S'12, M'20) received the B.Sc. and M.Sc. degrees (honors) in electrical engineering from the K. N. Toosi University of Technology, Tehran, Iran, in 2010 and 2012, respectively, and the Ph.D. degree (with distinctions) from the Friedrich-Alexander-University (FAU) of Erlangen-N\"urnberg, Erlangen, Germany, in 2019. In 2017, he was a Visiting Research Scholar with Stanford University, CA, USA. He is currently a Postdoctoral Research Fellow with the Department of Electrical and Computer Engineering at Princeton University. His research interests include wireless and molecular communications, Bayesian inference and learning, and multiuser information theory. 

Dr. Jamali has served as a member of the Technical Program Committee for several IEEE conferences and he is currently an Associate Editor of the \textsc{IEEE Communications Letters}, \textsc{IEEE Open Journal of the Communications Society}, and Elsevier Physical Communication Journal. He received several awards for his publications and research work including the Best Paper Awards from the IEEE International Conference on Communications in 2016, the ACM International Conference on Nanoscale Computing and Communication in 2019, the Asilomar Conference on Signals, Systems, and Computers in 2020, and the IEEE Wireless Communications and Networking Conference in 2021; the Doctoral Research Grant from the German Academic Exchange Service (DAAD) in 2017; the Goldener Igel Publication Award from the Telecommunications Laboratory (LNT), FAU, in 2018; the Best Ph.D. Thesis Presentation Award from the IEEE Wireless Communications and Networking Conference in 2018; the Best Journal Paper Award (Literaturpreis) from the German Information Technology Society (ITG) in 2020; and the Postdoctoral Research Fellowship by the German Research Foundation (DFG) in 2020.
\end{IEEEbiography}

\begin{IEEEbiography}
	[{\includegraphics[width=1in,height=1.25in,clip,keepaspectratio]{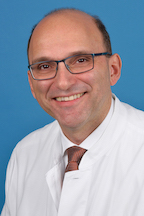}}]{Christoph Alexiou}
	    received his Ph.D.\ in 1995 from the TU-Munich, Medical school and 2002 he changed to the ENT-Department in Erlangen, Germany, where he performed his postdoctoral lecture qualification (Habilitation).
    He is working there as an assistant medical director in the clinic and leads the Section for Experimental Oncology and Nanomedicine (SEON).
    Since 2009 he owns the W3-Else Kröner-Fresenius-Foundation-Professorship for Nanomedicine at the University Hospital Erlangen.
    His research is addressing the emerging fields of Diagnosis, Treatment, Regenerative Medicine and Molecular Communication using magnetic nanoparticles.
    He received for his research several national and international renowned awards.
\end{IEEEbiography}

\begin{IEEEbiography}
[{\includegraphics[width=1in,height=1.25in,clip,keepaspectratio]{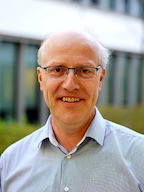}}]
{Georg Fischer}  (M’01–SM’08) was born in Wachtendonk, Germany, in 1965. He received the Diploma degree in electrical engineering with a focus on communications and microwave from RWTH Aachen University, Aachen, Germany, in 1992, and the Dr.Ing. degree in electrical engineering from the University of Paderborn, Paderborn, Germany, in 1997. From 1993 to 1996, he was a Research Assistant with the University of Paderborn, where he was involved in adaptive antenna array systems for mobile satellite communications. From 1996 to 2008, he performed research with Bell Laboratories, Lucent (later Alcatel-Lucent), where he focused on the RF and digital architecture of mobile communication base stations for global system for mobile communications (GSM), universal mobile telecommunications system (UMTS), and features for network coverage and capacity enhancements. In 2000, he became a Bell Laboratories Distinguished Member of Technical Staff (DMTS), and in 2001, he became a Bell Laboratories Consulting Member of Technical Staff (CMTS). He also acted as the Chairman of the European Telecommunications Standards Institute (ETSI) during the physical layer standardization of the GSM-EDGE system. From 2001 to 2007, he was a part-time Lecturer with the University of Erlangen-Nuremberg, Erlangen, Germany, during that time, he lectured on base station technology. Since 2008, he has been a Professor of electronics engineering with the University of Erlangen-Nuremberg.

He holds over 50 patents concerning microwave and communications technology. His research interests are transceiver design, analog/digital partitioning, converters, enhanced amplifier architectures, duplex filters, meta material structures, GaN transistor technology and circuit design, and RF micro-electromechanical systems (MEMS) with a specific emphasis on frequency agile, tunable, and reconfigurable RF systems for software-define radio (SDR) applications. His current research interests concentrate on medical electronics, such as using microwaves for detection of vital parameters. He is also a Senior Member of the IEEE Microwave Theory and Techniques (MTT), Antennas and Propagation (AP), Computer Society (COMSOC), and Vehicular Technology Society (VTC). He is also a member of VDE-ITG and the European Microwave Association (EUMA).
\end{IEEEbiography}

\begin{IEEEbiography}
	[{\includegraphics[width=1in,height=1.25in,clip,keepaspectratio]{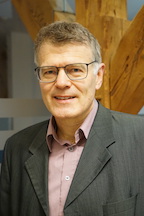}}]{Robert Schober} (S'98, M'01, SM'08, F'10) received the Diplom (Univ.) and
the Ph.D. degrees in electrical engineering from Friedrich-Alexander
University of Erlangen-Nuremberg (FAU), Germany, in 1997 and 2000,
respectively. From 2002 to 2011, he was a Professor and Canada Research
Chair at the University of British Columbia (UBC), Vancouver, Canada.
Since January 2012 he is an Alexander von Humboldt Professor and the Chair
for Digital Communication at FAU. His research interests fall into the
broad areas of Communication Theory, Wireless Communications, and
Statistical Signal Processing.
Robert received several awards for his work including the 2002 Heinz
Maier­ Leibnitz Award of the German Science Foundation (DFG), the 2004
Innovations Award of the Vodafone Foundation for Research in Mobile
Communications, a 2006 UBC Killam Research Prize, a 2007 Wilhelm Friedrich
Bessel Research Award of the Alexander von Humboldt Foundation, the 2008
Charles McDowell Award for Excellence in Research from UBC, a 2011
Alexander von Humboldt Professorship, a 2012 NSERC E.W.R. Stacie
Fellowship, and a 2017 Wireless Communications Recognition Award by the
IEEE Wireless Communications Technical Committee. Since 2017, he has been
listed as a Highly Cited Researcher by the Web of Science. Robert is a Fellow of the
Canadian Academy of Engineering, a Fellow of the Engineering Institute
of Canada, and a Member of the German National Academy of Science and Engineering.
\end{IEEEbiography}

\end{document}